\documentclass[10pt,journal,twocolumn,twoside]{IEEEtran} 

\usepackage[dvips]{epsfig}
\usepackage[dvips]{graphicx}
\usepackage[dvipsnames]{xcolor}
\usepackage{amsmath,amsfonts,bm,amssymb,psfrag,ifthen}
\allowdisplaybreaks[1]
\usepackage{algpseudocode}
\usepackage{algorithm}
\usepackage{algorithmicx}
\usepackage{ifthen}
\usepackage{setspace}
\usepackage{cite}
\usepackage{url}
\usepackage{longtable}
\usepackage{stfloats}  
\usepackage{graphics}
\usepackage{float}
\usepackage{array}
\usepackage{listings}

\floatname{algorithm}{Algorithm}

\begin{document}

\title{Robust Max-Min Fair Beamforming Design for Rate Splitting Multiple Access-aided Visible Light Communications}
  \author{Zhengqing~Qiu, Yijie~Mao,~\IEEEmembership{Member,~IEEE,} Shuai~Ma,~\IEEEmembership{Member,~IEEE,}
   Bruno~Clerckx,~\IEEEmembership{Fellow,~IEEE}
   \\
    \thanks{This work has been supported in part by the National Nature Science Foundation of China under Grant 62201347; and in part by Shanghai Sailing Program under Grant 22YF1428400. Shuai Ma was supported in part by the National Natural Science Foundation of China under Grant 62471270, and in part by Guangdong Basic and Applied Basic Research Foundation  under Grant  2024A1515030028.  \par
   Z. Qiu and Y. Mao are with the School of Information Science and Technology, ShanghaiTech University, Shanghai 201210, China (email: qiuzhq2023@shanghaitech.edu.cn, maoyj@shanghaitech.edu.cn). \par
   Shuai Ma is with Peng Cheng Laboratory, Shenzhen 518066, China (email: mash01@pcl.ac.cn) \par
   B. Clerckx is with the Department of Electrical \& Electronic Engineering, Imperial College London, London SW7 2AZ, U.K. (email: {b.clerckx}@imperial.ac.uk).
   \par (\textit{Corresponding Author: Yijie Mao})
   }
   
   }
    
\maketitle
\begin{abstract}

This paper addresses the robust beamforming design for rate splitting multiple access (RSMA)-aided visible light communication (VLC) networks with imperfect channel state information at the transmitter (CSIT). 
In particular, we first derive the theoretical lower bound for the channel capacity of RSMA-aided VLC networks.
Then we investigate the beamforming design to solve the max-min fairness (MMF) problem of RSMA-aided VLC networks under the practical optical power constraint and electrical power constraint while considering the practical imperfect CSIT scenario.
To address the problem, we propose a constrained-concave-convex programming (CCCP)-based beamforming design algorithm which exploits semidefinite relaxation (SDR) technique and a penalty method to deal with the rank-one constraint caused by SDR.
Numerical results show that the proposed robust beamforming design algorithm for RSMA-aided VLC network achieves a superior performance over the existing ones for space-division multiple access (SDMA) and non-orthogonal multiple access (NOMA).

\end{abstract}
\begin{IEEEkeywords}
Visible light communication,  rate splitting multiple access, imperfect channel state information,  max-min fairness, robust beamforming design.
\end{IEEEkeywords}

\IEEEpeerreviewmaketitle
\section{INTRODUCTION}

\IEEEPARstart{T}{he} field of wireless communication is undergoing significant changes with the continuous advancement of science and technology. 
In this era of rapid development, there is a pressing demand for faster, more secure and highly reliable means of communication. 
To address these requirements, the field of visible light communications (VLC) is widely studied as the forefront of technological innovation and exhibits significant potential for 6G and beyond.
In essence, VLC is a wireless communication technology which employs visible light frequency band for information transmission.
In a VLC system, the intensity or frequency of light is modulated to convert binary data into optical signals that can be received and decoded by an optical sensor.
VLC can offer several benefits, including an extensive unlicensed spectrum range from 380 - 790 THz, high spatial reuse, high energy efficiency, cost-effective front-ends, and intrinsic security.
The VLC technology is therefore considered as a promising technology for meeting the increasing demand for massive data in 6G indoor networks \cite{vlc_6g_1, vlc_6g_2, vlc_6g_3}.
It is appealing for various applications, such as smart homes, smart cities, and Internet of Things (IoTs).


While VLC offers significant advantages, it also faces certain challenges. 
The primary challenge for VLC is the limited modulation bandwidth of existing available light-emitting diodes (LEDs), which restricts both connectivity and spectral efficiency in VLC networks.
To overcome this challenge and enhance spectral efficiency, designing effective multi-access (MA) schemes is a promising research direction for VLC.
Conventional orthogonal multiple access (OMA) have been extensively researched in VLC networks. 
For example, optical orthogonal frequency-division multiple access (OFDMA) is investigated in \cite{Dang_WCSP_2012,Sung_OC_2015} and optical code-division multiple access (OCDMA) has been investigated in \cite{Medina_EL_2012}.
However, the number of active users served by the OMA schemes is limited by several factors such as the number of available orthogonal time, frequency and code resources.
To overcome these limitations, non-orthogonal multiple access (NOMA) has been proposed \cite{Ding_SPL_2014,Dai_comMag_2015,Ding_JSAC_2017,Marshoud_PTL_2016,mashuai_mobile}, which allows different users to share the same time or frequency resources.
Based on superposition coding (SC) at the transmitter and successive interference cancellation (SIC) at the receivers, 
power-domain NOMA has been shown to attain higher spectral efficiency and and improved connectivity compared to OMA \cite{Zhang_CL_2016,Yin_TCOM_2016,Guan_OE_2016}.
However, it is worth mentioning that the performance of NOMA highly depends on the channel conditions of the users.  
Its performance is degraded when users share similar channel strengths.
In addition, NOMA attains a high hardware complexity because of using multiple layers of SIC at the receivers.
In addition to NOMA, there is another well-known type of MA known as space division multiple access (SDMA). 
SDMA is commonly implemented through multi-user linear precoding (MU–LP).
It works as a well-established MA that is nowadays the basic principle behind numerous multi-antenna techniques in 4G and 5G.  
However, SDMA can not efficiently work in overloaded regimes, i.e., the number of users is more than the number of transmit antennas, and its performance is highly sensitive to channel accuracy, orthogonality and strengths among users.

In order to reduce the dependence of MA schemes on channel accuracy, orthogonality and channel strengths, a novel MA scheme named rate-splitting multiple access (RSMA) has been proposed based on linearly precoded rate-splitting (RS) at the transmitter and SIC at the receivers \cite{Clerckx_COMag_2016,rsma_fsfrt,Clerckx_2023_rsma_primer}.
Specifically, at the transmitter side, the message of each user is split into a common part and a private part.
The common parts are then encoded jointly into common streams, while the private parts are encoded independently into private streams.
By means of SC, all streams are simultaneously transmitted.
All users sequentially decode the intended common and private streams.
The primary benefit of RSMA lies in that it allows to adjust the message split and the power allocation among the common and private streams flexibly, thereby softly bridges the two extremes of fully treating interference as noise and fully decoding interference \cite{Mao_EURASIP_2018,Mao_TCOM_2019}.
Such powerful interference management capability has enabled RSMA to be a versatile multiple access scheme that subsumes SDMA and NOMA as special cases \cite{rsma_fsfrt,Mao_EURASIP_2018,Clerckx_2020_rsma_twouser_analysis}.
Meanwhile, the first-ever prototype of RSMA has been successfully realized and experiments have demonstrated the advantages and superiority of RSMA over SDMA and NOMA \cite{Lyu_rsma_surpass_sdmanoma}.

\subsection{Related Works}
In conventional radio frequency (RF) communications, RSMA has been extensively studied in both information theory and communication theory.
From an information theoretical perspective, RSMA has been proved to achieve the optimal spatial multiplexing gain in both underloaded and overloaded multi-antenna broadcast channels with imperfect channel state information at the transmitter (CSIT) \cite{Joudeh_2016_rsma_robustness,Piovano_2017_dof_rsma_imperfect,Mao_2021_rsma_overloaded_iot}. 
From a communication theory perspective, research has shown that RSMA surpasses existing multiple access schemes in spectral efficiency, energy efficiency, user fairness, Quality of Service (QoS) enhancements, and reliability in a wide range of network loads and user deployment \cite{rsma_fsfrt,Clerckx_2023_rsma_primer,Clerckx_2024_6g_survey}. 
It significantly enhances robustness against imperfect CSIT and user mobility \cite{Joudeh_2016_rsma_robustness,Dizdar_2021_rsma_robustness}, aligning with findings in information theory.
Motivated by advantages mentioned above, RSMA has been widely investigated in many emerging 6G applications, such as
massive multiple-input multiple-output (MIMO) \cite{Dai_TWC_2016,Anastasios_TVT_2017}, multigroup multicasting \cite{Joudeh_TWC_2017},
integrated communication and sensing \cite{ISAC_ref1,ISAC_ref2}, non-orthogonal unicast and multicast transmission \cite{Mao_TCOM_2019,Chen_TVT_2020}, cooperative transmission with user relaying \cite{Zhang_SPL_2019,Mao_TWC_2020}, etc.


RSMA has demonstrated significant advantages such as enhancing spectral efficiency, user fairness, and transmission robustness against CSIT imperfections in RF
communications. 
Inspired by these advantages, some studies have begun exploring the application of RSMA in VLC networks. However, its potential benefits in VLC are still in the early stages of investigation.
In \cite{Tao_ICC_2020,Naser_2022,Xing_2022}, RSMA is investigated in multi-cell VLC networks.
Similar to the existing studies in RF communications, these works utilized the classic Shannon capacity as the rate expression and proposed various beamforming design methods for different optimization targets such as maximizing spectral efficiency, minimizing the sum of mean squared error (MSE) among all users, and maximizing energy efficiency.
These works all show that RSMA is superior to existing SDMA and NOMA strategies in VLC networks.

Although these works have certain merits, the limitations cannot be ignored:
\begin{itemize}
    \item The first limitation is that most of existing works on RSMA VLC
\cite{Tao_ICC_2020,Naser_2022} only consider applying the classic Shannon formula for problem formulation.
Nevertheless, the Shannon formula, derived based on Gaussian input distributions, is not appropriately applicable to VLC networks. 
This is due to the unique characteristics of optical channels, the non-negativity constraint on transmitted signals, and the different power constraints inherent in VLC systems. 
Specifically, practical illumination demands and user eye safety must be considered in VLC networks. 
Additionally, the use of intensity modulation and direct detection (IM/DD) in VLC networks ensures that the transmit signal is both real and non-negative.
In \cite{Smith,Chan_1}, the distribution of VLC channels is demonstrated to be discrete over a finite set of points, and the exact capacity cannot be expressed in closed form.
Therefore, the classic Shannon capacity formula usually used in RF systems cannot accurately evaluate the achievable rate of VLC networks. 
Specialized capacity formulations and analyses that account for these unique characteristics are necessary for VLC networks.
Unfortunately, the capacity of the single-input single-output (SISO) VLC channel remains unknown.
Many works turned to investigate upper and lower bounds on the channel capacity of VLC networks.
However, most of existing works \cite{Xing_2022,2015_pham,2024_guo,2013_wang,2009_lapidoth} only consider peak optical power or average optical power while ignoring electrical power constraints when deriving the distribution of the input to obtain the upper and lower bounds on the channel capacity.
\item Secondly, existing works on RSMA-aided VLC \cite{Torky,Naser_2022,Tao_ICC_2020,Naser_Shimaa_2020,Naser_Shimaa_A_2021} consider only perfect CSIT.
However, in practice, the estimation errors of CSI are inevitable at the transmitter due to quantization errors and user mobility. 
As RSMA has demonstrated significant robustness towards CSIT imperfections in RF communications, an in-depth study of the robustness of RSMA-aided VLC networks with imperfect CSIT is warranted.
\end{itemize}



\subsection{Contributions}
Motivated by the above two major limitations of existing works, this paper focuses on addressing two fundamental challenges in RSMA-aided VLC networks: 1) deriving lower bounds of achievable rates; 2) designing robust beamforers at the transmitter under imperfect CSIT scenarios.
The primary contributions of this paper are summarized as:
 \begin{itemize}
    \item 
    Considering the unique characteristics of VLC networks, including the distinct signal distribution and practical constraints on average and peak optical power and average electric power, we derive closed-form expressions for the lower bounds of the achievable rate based on the entropy power inequality and entropy inequality.
    To the best of our knowledge, this is the first work that derives theoretical lower bounds for the achievable rates of RSMA-aided VLC networks. 
    Such closed-form rate expressions are essential since it is a prerequisite for the application of RSMA in VLC networks.

    \item 
    Based on the practical imperfect CSIT scenario, the consequent imperfect SIC, as well as the derived rate expression, we formulate and investigate a robust beamforming design problem to maximize the worst-case achievable rate among users, namely the max-min fairness (MMF) rate.
    To the best of our knowledge, this is the first work that investigates the robust beamforming design of RSMA-aided
    VLC networks with imperfect CSIT.

    \item 
    The derived achievable rate expression is highly nonconvex, and incorporating imperfect CSIT complicates the problem further compared to existing ones. To address these challenges, we propose an efficient optimization algorithm.
    Specifically, we first apply semidefinite relaxation (SDR) and constrained-concave-convex programming (CCCP) methods to transform the original non-convex problem. 
    Then, we apply a penalty method to address the rand-one constraint. 
    The problem is then solved by consecutively solving a series of convex subproblems.
    Numerical results demonstrate the explicit MMF rate gain of RSMA over SDMA and NOMA, as well as its enhanced robustness to imperfect CSIT in VLC networks.

  \end{itemize}
  \begin{figure*}[ht]
      \centering
	\includegraphics[width=14.11cm]{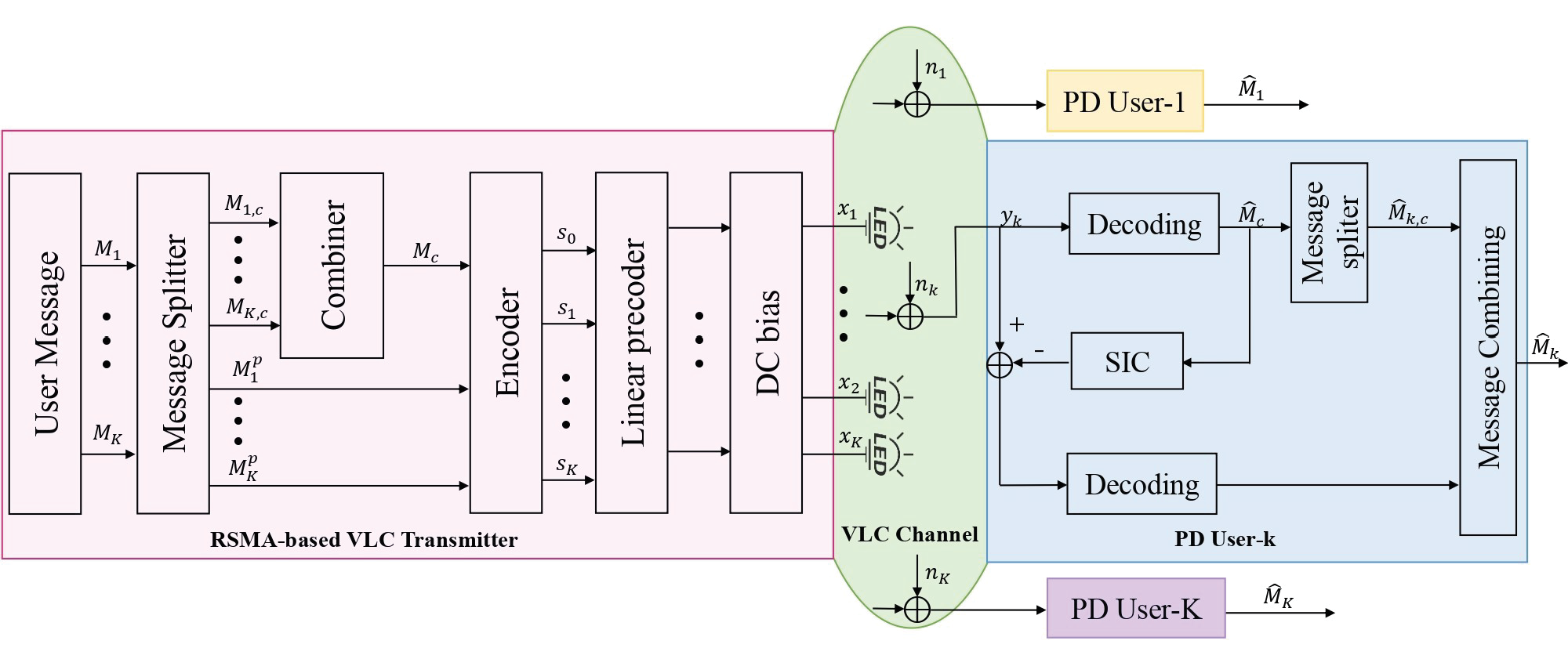}
 \caption{The considered downlink RSMA-aided multi-user MISO VLC network.}
  \label{system_model} 
\end{figure*}
  \par
  \subsection{Organizations and Notations}
 { 
\par \textit{Organizations}:   The rest of this paper is structured as follows. 
  In Section \uppercase\expandafter{\romannumeral2},
  we present four subsections that respectively introduce the system model of the downlink RSMA-aided VLC network, the imperfect CSIT model, the derivation of the lower bounds of the achievable rates, and the formulation
of the MMF rate problem.
  In Section \uppercase\expandafter{\romannumeral3}, 
  we present the proposed robust beamforming design algorithm in four subsections, each covering the semidefinite relaxation, constraint transformation, CCCP, and the penalty method used in our algorithm.
  In Section \uppercase\expandafter{\romannumeral4}, numerical results are illustrated and discussed. 
  Section \uppercase\expandafter{\romannumeral5} draws the conclusions.
  }
  \par \textit{Notations}: Vectors and matrices are denoted by boldfaced lowercase and uppercase letters, respectively.
$\left| \cdot \right|$, ${\mathbb{E}}\left\{ \cdot \right\}$,
${\left(  \cdot  \right)^{{T}}}$, $\left\|  \cdot  \right\|$, ${\mathrm{Rank} \left( \cdot \right)}$, and ${\mathrm{Tr}}\left( \cdot \right)$ represent absolute value,  the expectation, transpose, Frobenius norm, rank and trace, respectively.
${\mathbf{1}}_N$ represents an $N \times 1$ vector where all elements are equal to $1$. 
${\mathbf{I}}_N$ represents an ${N} \times {N}$ identity matrix.
${\mathbb{R}^{m \times n}}$ represents an $m$ by $n$ dimensional real space. 


\section{SYSTEM MODEL AND PROBLEM FORMULATION}

\subsection{ Transmit Signal }\label{Transmit Signal}

As depicted in Fig. \ref{system_model}, we consider a downlink RSMA-aided multi-user multiple-input single-output (MISO) VLC network here.
The VLC transmitter is equipped with $N$ LEDs indexed by $\mathcal{N}=\{1,\ldots,N\}$, and it simultaneously serves $K$ single photodiode (PD) users indexed by $\mathcal{K}=\{1,\ldots,K\}$. 
The entire transmission process is facilitated by 1-layer RSMA \cite{rsma_fsfrt}.
Specifically, at the VLC transmitter, the message $M_k$ for user $k$ is split into two messages, namely, a common message $M_{k,c}$ and a private message $M_{k,p},\forall k \in {\mathcal{K}}$. 
The common parts of all $K$ users $\left\{ {{{M}_{k,c}}} \right\}_{k = 1}^K$ are combined into one common message ${M_c}$, i.e., $M_c = \left\{ M_{1,c}, \ldots ,M_{K,c} \right\}$. 
By employing a shared codebook, the common message $M_c$ is encoded into a common stream $s_0$, facilitating all users to decode it.
Meanwhile, by leveraging a private codebook, each private message $M_{k,p}$ is independently encoded into a private stream $s_k$, exclusively decodable by user $k$.

Due to the LED characteristics, these $K+1$ signals $\left\{ {{s_i}} \right\}_{i = 0}^K$ should meet the peak amplitude requirement $\left| {{s_i}} \right| \le {A_i}$.
The mean and variance of $s_i$ respectively follow $\mathbb{E} \left \{ s_i \right \} =0 $ and $\mathbb{E} \left \{ s_i^2 \right \} =\varepsilon _i$, $\forall i \in {\mathcal{I}} = \left\{ {0,1, \ldots ,K} \right\}$ \cite{S_Ma_2017}.
Each stream $s_i$ is linearly precoded by the beamforming vector ${{\mathbf{p}}_i} = {\left[ {{p_{i,1}}, \ldots ,{p_{i,N}}} \right]^T} \in {{\mathbb{R}}^{N \times 1}}$, $  \forall i \in \mathcal{I}$.
The resulting transmit signal is given as 
\begin{align}
{\mathbf{x}} = \sum\limits_{i = 0}^K {{{\mathbf{p}}_i}{s_i} + } {\mathbf{b}},
\end{align}
where  ${\mathbf{b}} = {\left[ {b, \ldots ,b} \right]^T} \in {{\mathbb{R}}^{N \times 1}}$ represents the direct current (DC) bias vector to assure the transmit signal being non-negative, i.e., $\mathbf{x}\geq\mathbf{0}$.
$b \ge {\mathrm{0}}$ is the DC bias for each LED \cite{Pathak}. 
Given the VLC features, the beamforming vectors $\{\mathbf{p}_i\}_{i=0}^K$ should satisfy \cite{Q_Gao}
\begin{align}
\sum\limits_{i = 0}^K {{A_i}\left| {{p_{i,n}}} \right|}  \le b, \forall n \in \mathcal{N}.
\end{align}

Besides the aforementioned non-negative intensity constraint $\mathbf{x}\geq\mathbf{0}$, the illumination requirements \cite{vlc_linear_constraint_ref} should be met in a VLC network. 
Specifically, we define the maximum permissible current of LEDs as $I_{\mathrm{H}}$ and the minimum permissible current of LEDs as $I_{\mathrm{L}}$. 
Then, the beamforming vectors $\{\mathbf{p}_i\}_{i=0}^K$ should also satisfy
 \begin{align}
{I_{\mathrm{L}}} \le \sum\limits_{i = 0}^K {{A_i}{\mathbf{p}}_i^T{{\mathbf{e}}_n} + b \le {I_{\mathrm{H}}},\forall n \in {\mathcal{N}}},
\end{align}
where ${\mathbf{e}}_n$ is an $N \times 1 $ unit vector with the $n$th element equal to 1 and all other elements equal to $0$.

\subsection{ Channel State Information }\label{Channel State Information}
Light propagation which travels from each LED to the individual user receiver typically comprises two components \cite{T_Q_Wang_2013}: 
a line-of-sight (LOS) component that is directly transmitted from the LED to the receiver, and a non-LOS (NLOS) diffuse reflection component that is propagated through reflection. 
Previous works \cite{2009_los_stronger1,2009_los_stronger2} have shown that the LOS component is generally far stronger than the diffuse component.  The optical wireless channel is therefore primarily influenced by the LOS link in VLC network, and the NLOS diffuse links can be safely neglected \cite{T_Fath_2013,T_Q_Wang_2013}. 
Therefore, in this paper, we merely consider the LOS component and defer the analysis of multipath transmissions to future studies.
In this work, in order to characterise the VLC channel between the LEDs and PD users, we follow the Lambertian emission model \cite{J_Kahn_1997}.
Specifically, we denote the channel vector between all LEDs and user $k$ as ${{\mathbf{h}}_k}   = {\left[ {{h_{k,1}}, \ldots ,{h_{k,N}}} \right]^T} \in \mathbb{R}^{N \times 1}$
, where the channel $h_{k,n}$ between the $n$th LED and user $k$ is given by
\begin{align} \label{channel_cal}
{h_{k,n}} = \frac{{\left( {l + 1} \right){\theta  _l}{\theta  _c}{A_{\mathrm{r}}}}}{{2\pi d_{k,n}^2}}{\cos ^l} \left( {{\phi _n}} \right)\cos \left( {{\varphi _k}} \right)\mathbb{I}\left ( \varphi _k \right )   ,
\end{align}
where $\theta _l$ and $\theta _c$ represent the conversion efficiency from light to current and the opposite, respectively; $l = - \frac{\ln{2} }{\ln{ \left ( \cos \left ( \phi _{1/2} \right )  \right ) }} $ represents the Lambertian order with the half-power angle ${{\phi _{{1 \mathord{\left/{\vphantom {1 2}} \right.\kern-\nulldelimiterspace} 2}}}}$; $\phi_n$ and ${{\varphi _k}}$ represent the radiance angle and incidence angle, respectively; ${d_{k,n}}$ represents the distance between user $k$ and the $n$th LED; $\mathbb{I}\left ( \varphi _k \right )$ represents an indicator function. 
Denote $\phi_{FOV}$ as the field of view (FOV) of each PD. 
When ${{\varphi _k}}$ satisfies ${\left| {{\varphi _k}} \right| \le {\varphi _{{\mathrm{FOV}}}}}$, $\mathbb{I} \left( {{\varphi _k}} \right) = 1$, otherwise $\mathbb{I} \left( {{\varphi _k}} \right) = 0$; ${{A_{{\mathrm{r}}}}}$ represents the effective physical area of each PD, which is given by
\begin{align}
{A_{\mathrm{r}}} = \frac{{n_r^2}}{{{{\sin }^2}\left( {{\varphi _{{\mathrm{FOV}}}}} \right)}}{A_{{\mathrm{PD}}}},
\end{align}
where $n_r$ represents the refractive index and $A_{{\mathrm{PD}}}$ represents the PD area.

In this work, we consider a practical imperfect CSIT scenario which takes users' activity area into consideration.
Specifically, we make the assumption that the locations of all users are bounded by their respective activity area during each data transmission block.
The activity area of each user, also known as the uncertainty region, is assumed to be on the horizontal plane, that is, the plane consisting of the $x$ and $y$ axes, and the vertical height of the horizontal plane is assumed to be fixed.
Moreover, the uncertainty region of each user is assumed to be within a circle with a fixed center $(x_k,y_k,z_k)$ and a radius of $r_k$, i.e., 
\begin{align}
\mathcal{R}_k  =  \left \{ (u_k,v_k)  | 0 \leq \left ( u_k - x_k \right )^2 +\left ( v_k - y_k \right )^2 \leq r_k^2\right \}. 
\end{align}

\begin{figure}[ht]
      \centering
	\includegraphics[width=8.11cm]{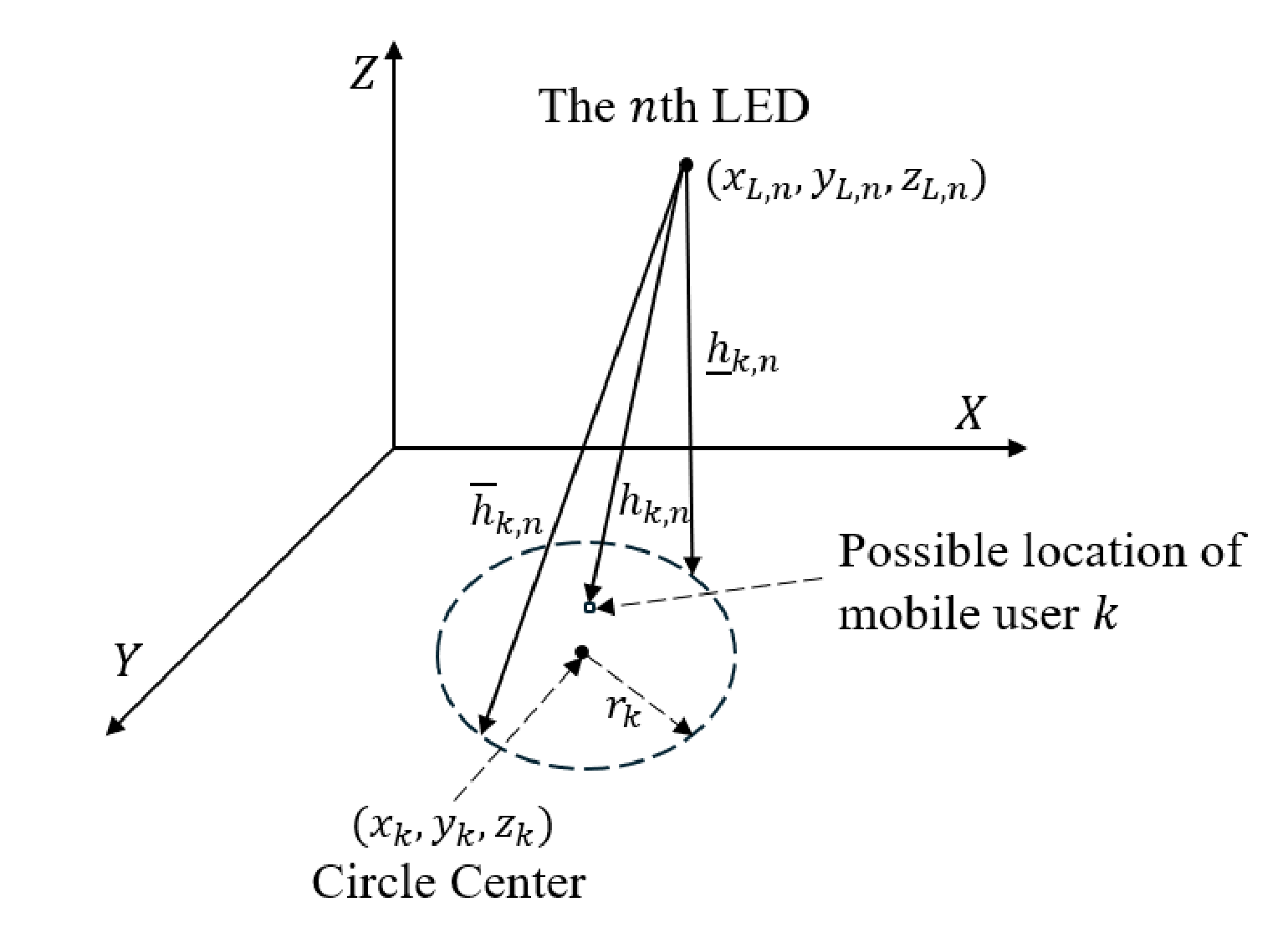}
\caption[]{The geometry of mobile users.}
  \label{mobility}
\end{figure}

The geometry of channel model of user $k$ is depicted in Fig. \ref{mobility}.
 
Let $(x_{L,n},y_{L,n},z_{L,n})$ denote the location of the $n$th LED. 
The distance between the $n$th LED and user $k$ is bounded, i.e., $d_{k,n}^{\mathrm{min}} \leq d_{k,n} \leq d_{k,n}^{\mathrm{max}} $, where
$d_{k,n}^{\mathrm{min}}$ and $d_{k,n}^{\mathrm{max}}$ represent the minimum distance and the maximum distance between the $n$th LED and user $k$, respectively.
Their values can be calculated as follows 
\begin{subequations} \label{bounded_dist}  
\begin{align}
d_{k,n}^{\mathrm{min}} =
\begin{cases}z_{L,n}-z_k,&\text{if} \left(x_{L,n},y_{L,n}\right)\in\mathcal{R}_k\\
 \sqrt{\left(d_{n,k}^{x,y}-r_k\right)^2+\left(z_{L,n}-z_k\right)^2} ,&\text{otherwise}
\end{cases}
\end{align}
\begin{align}
d_{k,n}^{\mathrm{max}}=
\sqrt{ \left(d_{k,n}^{x,y}+r_k\right)^2+\left(z_{L,n}-z_k\right)^2},
\end{align}
\end{subequations}
where $d_{k,n}^{x,y} = \sqrt{\left ( x_{L,n}-x_k \right )^2 + \left ( y_{L,n}-y_k \right )^2  } $ represents the distance between the $n$th LED and user $k$ in the $x-y$ plane.

In order to take users' activity area into consideration, we adopt an imperfect CSIT model with bounded errors \cite{Liu_Xiaodong2020,Ma_Shuai_2016}.
Specifically, we denote estimated CSIT at the VLC transmitter for user $k$ as $\mathbf{\hat h}_k$. 
The imperfect CSIT is modeled as  
\begin{align}\label{imperfect_CSI_model1}
{{\mathbf{h}}_k} = {{{\mathbf{\hat h}}}_k} + \Delta \mathbf{h}_k,
\end{align}
where ${{{\mathbf{\hat h}}}_k} = {\left[ {{{\hat h}_{k,1}}, \ldots ,{{\hat h}_{k,N}}} \right]^T}\in {{\mathbb{R}}^{N \times 1}}$ represents the estimated CSIT vector for user $k$, and $\Delta {{\mathbf{h}}_k} = {\left[ {\Delta {h_{k,1}}, \ldots ,\Delta {h_{k,N}}} \right]^T}\in {{\mathbb{R}}^{N \times 1}}$ represents the CSIT estimation errors vector for user $k$.
In indoor VLC systems, one primary source of estimation error is user mobility.
Therefore, the active area of users affects the CSIT estimation error, indicating that the bound of this CSIT estimation error is directly influenced by the radius of the users' activity area.
Specifically, $\hat h_{k,n} = \frac{1}{2} \left ( \overline{h}_{k,n} + \underline{h}_{k,n}   \right ) $ and $ \left | \Delta h_{k,n} \right |  \leq \frac{1}{2} \left ( \overline{h}_{k,n} - \underline{h}_{k,n}   \right )$, where $\overline{h}_{k,n}$ and $\underline{h}_{k,n}$ are the upper and lower bounds of the real channel ${h}_{k,n}$ \cite{mashuai_mobile}. 
$\overline{h}_{k,n}$ and $\underline{h}_{k,n}$ can be directly calculated based on $d_{k,n}^{\mathrm{max}}$ and $d_{k,n}^{\mathrm{min}}$ in (\ref{bounded_dist}).
Following \cite{Ma_Shuai_2016}, the CSI estimation errors are bounded within the set :  
\begin{align}
{{\mathcal H}_k}  = \left\{ {\Delta {{\mathbf{h}}_k}\left| {\Delta {\mathbf{h}}_k^T\Delta {{\mathbf{h}}_k} \le {v_k}} \right.} \right\}\label{imperfect_CSI_model2},
\end{align}
where $v_k$ represents uncertainty region determined by the user's activity area. 
Mathematically, $v_k = \frac{1}{4} \left ( \overline{\mathbf{h}}_{k} - \underline{\mathbf{h}}_{k}   \right )^T \left( \overline{\mathbf{h}}_{k} - \underline{\mathbf{h}}_{k}   \right )$, where $\overline{\mathbf{h}}_{k} = \left[ \overline{h}_{k,1}, \overline{h}_{k,2},\ldots ,\overline{h}_{k,N} \right]^T$ and $\underline{\mathbf{h}}_{k} = \left[ \underline{h}_{k,1}, \underline{h}_{k,2},\ldots ,\underline{h}_{k,N} \right]^T$, $\forall k \in \mathcal{K}$.
When ${v_k} = 0$, CSI is perfect known at the transmitter, i.e., $\Delta {{\mathbf{h}}_k} = {\mathbf{0}}$.


\subsection{ Received Signal and Achievable Rates}
Based on the transmit signal and the imperfect CSIT model respectively specified in the previous two subsections, the receive signal at user $k$ is obtained as 
\begin{align}\label{receive_signal}
y_k = \underbrace{\left ( {\mathbf{\hat h}}_k^T + \Delta \mathbf{h}_k^T \right ) \mathbf{p}_0 s_0 }_{\mathrm{common \ stream}} + 
& \underbrace{ \sum_{i=1}^{K} \left ( {\mathbf{\hat h}}_k^T + \Delta \mathbf{h}_k^T \right ) \mathbf{p}_i s_i }_{\mathrm{private \ stream}} \nonumber \\
 + \underbrace{\left ( {\mathbf{\hat h}}_k^T + \Delta \mathbf{h}_k^T \right ) \mathbf{b} }_{\mathrm{DC\ component}} & + n_k ,
\end{align}
where $n_k$ represents Gaussian noise received at user $k$ with zero mean and variance $\sigma_k^2$.
The DC component $\left ( {\mathbf{\hat h}}_k^T + \Delta \mathbf{h}_k^T \right ) \mathbf{b}$ carries no message and it can be removed through the capacitor at user $k$ \cite{Tao_ICC_2020}.

By treating all private streams as noise, user $k$ first decodes the common stream ${s_0}$. 
Following \cite{S_Ma_2017}, the achievable rate of the common stream $s_0$ at user $k$ is given by
\begin{subequations}\label{ro_common_rate}
\begin{align}
&{R_{k,{\mathrm{c}}}} = \mathop {\max }\limits_{\left\{ {{f_{{s_k}}}({s_k})} \right\}_{k = 0}^K} {\mathrm{I}}\left( {{s_{\mathrm{0}}};{y_k}} \right)\\
&= \mathop {\max }\limits_{\left\{ {{f_{{s_k}}}\left( {{s_k}} \right)} \right\}_{k = 0}^K} H\left( {{y_k}} \right) - H\left( {{y_k}|{s_0}} \right)\\
&= \mathop {\max }\limits_{\left\{ {{f_{s_i}}\left( {{s_i}} \right)} \right\}_{i = 0}^K} H\left( {\sum\limits_{i = 0}^K {\left( {{\mathbf{\hat h}}_k^T + \Delta {\mathbf{h}}_k^T} \right){{\mathbf{p}}_i}{s_i}}  + {n_k}} \right)\nonumber\\ 
&- H\left( {\sum\limits_{j = 1}^K {\left( {{\mathbf{\hat h}}_k^T + \Delta {\mathbf{h}}_k^T} \right){{\mathbf{p}}_j}{s_j}}  + {n_k}} \right) \label{common_derivation_9c}
\end{align}
\end{subequations}
where ${\mathrm{I}}\left( {X;Y} \right)$ represents the mutual information between input $X$ and output $Y$, $f_{s_i}\left( {s_i} \right)$ represents the PDF of the signal $s_i$ and $H \left( x \right)  =  - \int {f\left( x \right)} {\log _2}f\left( x \right){\mathrm{d}}x$ represents the entropy of random variable $x$. 
Equation \eqref{common_derivation_9c} is intractable.
To further derive a tractable $R_{k,c}$, we further apply the entropy power inequality and entropy inequality respectively to the first and the second term of \eqref{common_derivation_9c}. 
Specifically, for independent variables $X$ and $Y$, the entropy power inequality is defined as 
\begin{align} \label{ei_1}
2^{2H\left ( X+Y \right )  } \geq 2^{2H(X)  } +2^{2H(Y) }.
\end{align}

For a random variable with variance $\mathrm{Var} (Z)$, the following entropy inequality holds :
\begin{align} \label{ei_2}
H(Z) \leq \frac{1}{2} \log_{2}{\left ( 2\pi e\mathrm{Var} \left ( Z \right )  \right )}.
\end{align}

Based on \eqref{ei_1} and \eqref{ei_2}, we then obtain a lower bound of \eqref{common_derivation_9c}, which is given as:
\begin{align}
\eqref{common_derivation_9c} \geq & \mathop {\max }\limits_{\left\{ {{f_{{s_i}}}\left( {{s_i}} \right)} \right\}_{i = 0}^K} \frac{1}{2}{\log _2}\left( {\sum\limits_{i = 0}^K {{2^{2H\left( {\left( {{\mathbf{\hat h}}_k^T + \Delta {\mathbf{h}}_k^T} \right){{\mathbf{p}}_i}{s_i}} \right)}}}  + {2^{2H\left( {{n_k}} \right)}}} \right) \nonumber\\
 - &\frac{1}{2}{\log _2}2\pi {\mathrm{e}}\left( {\sum\limits_{j = 1}^K {{{\left| {\left( {{\mathbf{\hat h}}_k^T + \Delta {\mathbf{h}}_k^T} \right){{\mathbf{p}}_j}} \right|}^2}{\varepsilon _j}}  + \sigma _k^2} \right) 
  =  \bar{R} _{k,c}\label{common_derivation_9d} 
\end{align}

Following Theorem $1$ in \cite{S_Ma_2017}, while comprehensively considering   three practical power constraints of the input, namely the peak optical power, the average optical power, and the electrical power constraints, we obtain a novel rate lower bound for each stream $s_i$ assuming the input follows the following distribution:
\begin{align}\label{abg_distribution}
f_{s_i} \left ( s_i \right )  = \begin{cases}
e^{-1-\alpha _i - {\beta _i}{s_i}- \gamma_{i} s_i^2}, & - {A_i} \le {s_i} \le {A_i};\\
0,  & \mathrm{otherwise} ,
\end{cases}  
\end{align}
where $\alpha_i $, $\beta_i $ and $\gamma_i$ are obtained by solving the following equations 
\begin{align}\label{ABG_para}
\begin{cases}
{T}\left( {{A_i}} \right) - {T}\left( { - {A_i}} \right) = {e^{1 + {\alpha _i}}} \\
 {\beta _i}\left( {{e^{{A_i}\left( {{\beta _i} - {\gamma _i}{A_i}} \right)}} - {e^{ - {A_i}\left( {{\beta _i} + {\gamma _i}{A_i}} \right)}} - {e^{1 + {\alpha _i}}}} \right) = 0\\
{e^{{A_i}\left( {{\beta _i} - {\gamma _i}{A_i}} \right)}}\left( {\left( {{\beta _i} - 2{\gamma _i}{A_i}} \right){e^{ - 2{A_i}{\beta _i}}} - {\beta _i} - 2{\gamma _i}{A_i}} \right)\\
 = 4\gamma _i^2{\varepsilon _i}{e^{1 + {\alpha _i}}}-\left( {\beta _i^2 + 2{\gamma _i}} \right){e^{1 + {\alpha _i}}} 
\end{cases}
\end{align}

Here, the function $T(x)$ and the error function $\mathrm{erf}(x)$ are given as
\begin{subequations}\label{t_erf}
\begin{align}
& T\left( x \right) = \sqrt \pi  \frac{{{e^{\frac{{\beta _i^2}}{{4{\gamma _i}}}}}{\mathrm{erf}}\left( {\frac{{{\beta _i} + 2{\gamma _i}x}}{{2\sqrt {{\gamma _i}} }}} \right)}}{{2\sqrt {{\gamma _i}} }},\\
& \mathrm{erf}\left( x \right) = \frac{2}{\sqrt \pi}  \int_0^x {\exp \left( { - {t^2}} \right)} dt.
\end{align}
\end{subequations}

Following the above distribution, we finally transform the lower bound of achievable rate for decoding the common stream $s_0$ at user $k$ as 
\begin{align}
\bar{R} _{k,c} =  \frac{1}{2}{\log _2}\frac{{2\pi \sigma _k^2 + \sum\limits_{i = 0}^K {{{\left| {\left( {{\mathbf{\hat h}}_k^T + \Delta {\mathbf{h}}_k^T} \right){{\mathbf{p}}_i}} \right|}^2}{ \tau_i}} }}{{2\pi \sigma _k^2 + 2\pi \sum\limits_{j = 1}^K {{{\left| {\left( {{\mathbf{\hat h}}_k^T + \Delta {\mathbf{h}}_k^T} \right){{\mathbf{p}}_j}} \right|}^2}{\varepsilon _j}} }} ,\label{common_rate_lower_bound}
\end{align}
where ${\tau _i}   = {e^{1 + 2\left( {{\alpha _i} + {\gamma _i}{\varepsilon _i}} \right)}}$. \par
After decoding the common stream $s_0$ successfully, $s_0$ is removed via SIC, and user $k$ then decodes the desired private stream $s_k$. 
It is worth noting that, under an imperfect CSIT scenario, user $k$ is not able to fully eliminate the common stream $s_0$ via the SIC procedure.
This results in residual interference. 
Therefore, the residual received signal at user $k$ after SIC is given as
\begin{align}
y_{k}^{\mathrm{SIC}} & = \underbrace{\left ( {\mathbf{\hat h}}_k^T + \Delta \mathbf{h}_k^T \right )\mathbf{p}_{k}s_{k}}_{\text{desired private stream}}+\underbrace{\Delta\mathbf{h}_{k}^{T}\mathbf{p}_{0}s_{0}}_{\text{residual common stream}} \nonumber \\
&+\underbrace{\sum_{i=1,i\neq k}^K\left ( {\mathbf{\hat h}}_k^T + \Delta \mathbf{h}_k^T \right )\mathbf{p}_is_i}_{\text{interference from other private streams}}+\left ( {\mathbf{\hat h}}_k^T + \Delta \mathbf{h}_k^T \right )\mathbf{b}+n_k,
\end{align}
where the term $\Delta {\mathbf{h}}_k^T{{\mathbf{p}}_0}{s_0}$ represents the residual interference from the common stream caused by channel estimation errors.
Following the same derivation procedure as in \eqref{ro_common_rate} and \eqref{common_derivation_9d}, we then obtain the achievable rate for decoding the desired private stream $s_k$ at user $k$, which is given by
\begin{subequations} \label{ro_private_rate}
\begin{align}
&{R_{k,{\mathrm{p}}}} = \mathop {\max }\limits_{\left\{ {{f_{s_k}}\left( {{s_k}} \right)} \right\}_{k = 0}^K} {\mathrm{I}}\left( {{s_k}; y_k^{{\mathrm{SIC}}}} \right)\label{ro_private_rate_a}\\
&= \mathop {\max }\limits_{\left\{ {{f_{s_k}}\left( {{s_k}} \right)} \right\}_{k = 0}^K} H\left( {y_k^{{\mathrm{SIC}}}} \right) - H\left( {y_k^{{\mathrm{SIC}}}|{s_k}} \right)\label{ro_private_rate_b}\\
&= \mathop {\max }\limits_{\left\{ {{f_{s_i}}\left( {{s_i}} \right)} \right\}_{i = 0}^K} H\left( {\Delta {\mathbf{h}}_k^T{{\mathbf{p}}_0}{s_0} + \sum\limits_{i = 1}^K {{\mathbf{h}}_k^T{{\mathbf{p}}_i}{s_i}}  + {n_k}} \right)\nonumber\\
&- H\left( {\Delta {\mathbf{h}}_k^T{{\mathbf{p}}_0}{s_0} + \sum\limits_{j = 1,j \ne k}^K {{\mathbf{h}}_k^T{{\mathbf{p}}_j}{s_j} + {n_k}} } \right)\label{ro_private_rate_c}\\
&\ge\mathop {\max }\limits_{\left\{ {{f_i}\left( {{s_i}} \right)} \right\}_{i = 0}^K} \frac{1}{2}{\log _2}\left( {\sum\limits_{i = 1}^K {{2^{2H\left( {{{\hat y}_{k,i}}} \right)}}} }\right)- \frac{1}{2}{\log _2}2\pi e\sigma _k^2\nonumber\\
&- \frac{1}{2}{\log _2}2\pi e\left( {{{\left| {\Delta {\mathbf{h}}_k^T{{\mathbf{p}}_0}} \right|}^2}{\varepsilon _0} + \sum\limits_{j = 1,j \ne k}^K {{{\left| {{\mathbf{h}}_k^T{{\mathbf{p}}_j}} \right|}^2}{\varepsilon _j} 
} } \right)  =  \bar{R} _{k,p},\label{ro_private_rate_d}
\end{align}
\end{subequations}
where ${\hat y}_{k,i}  = \Delta {\mathbf{h}}_k^T{{\mathbf{p}}_0}{s_0} + {\mathbf{h}}_k^T{{\mathbf{p}}_i}{s_i} + {n_k}$. 

Then, following the same distribution in \eqref{abg_distribution},\eqref{ABG_para} and \eqref{t_erf}, we then obtain the lower bound of the achievable rate for decoding the desired private stream $s_k$ at user $k$ as
\begin{align}
\bar{R} _{k,{\mathrm{p}}} 
= \frac{1}{2}{\log _2}\frac{{2\pi \sigma _k^2 +\left | \Delta \mathbf{h}_k^T \mathbf{p}_0 \right |^2 \tau_0  +\sum\limits_{i = 1}^K {{{\left| {{\mathbf{h}}_k^T{{\mathbf{p}}_i}} \right|}^2}{\tau _i}} }}
{{2\pi \left ( \sigma _k^2 + \left | \Delta \mathbf{h}_k^T \mathbf{p}_0 \right |^2 \varepsilon _0  + \sum\limits_{j = 1,j \ne k}^K {{{\left| {{\mathbf{h}}_k^T{{\mathbf{p}}_j}} \right|}^2}{\varepsilon _j}}\right )    }}.\label{private_rate_lower_bound}
\end{align}
In order to make sure that all users can successfully decode ${s_0}$, the lower bound of achievable rate for decoding ${s_0}$ should not exceed $\bar{R} _{\mathrm{c}} = \min_{k \in \mathcal{K} } \left\{ \bar{R} _{k,{\mathrm{c}}}  \right\}$ \cite{Clerckx_Bruno_2021,Mao_EURASIP_2018}. 
While $\bar{R} _{\mathrm{c}}$ is shared by all users for the transmission of $s_0$, we have $\sum\nolimits_{k = 1}^K {{c_k} = {\bar{R} _{\mathrm{c}}}}$, where ${c_k}$ is the portion of $\bar{R} _{\mathrm{c}}$ allocated to user $k$ \cite{Mao_EURASIP_2018}.
Therefore, following the principle of $1$-layer RSMA, we can calculate the total lower bound of achievable rate ${\bar{R} _k}$ at user $k$ as
\begin{align}\label{total_rate}
\bar{R} _{k}=c_{k} + \bar{R} _{k,{\mathrm{p}}}, ~\forall k \in \mathcal{K}.
\end{align}
\subsection{ Problem Formulation}
In this work, we primarily concentrate on the robust beamforming design for RSMA-aided VLC networks with imperfect CSI. 
Specifically, our goal is to maximize the worst-case rate among users subject to a series of constraints at the VLC transmitter.
The worst-case rate (also known as MMF rate) among users is expressed as $\min_{k\in \mathcal{K} } \left \{  \bar{R} _k \right \} $.
For the RSMA-aided VLC networks, we can formulate MMF rate maximization problem as :
\begin{subequations}\label{problem_ini}
\begin{align}
 \mathbf{\mathcal{P}_0 :}   \ \ \ & \mathop {\max }\limits_{\left\{ {{{\mathbf{p}}_i}} \right\}_{i = 0}^K,\left\{ {{c_k}} \right\}_{k = 1}^K} \; \ \min_{k\in \mathcal{K} } \left \{  \bar{R} _k \right \} \label{problem_ini_ob}\\
{\mathrm{s}}.{\mathrm{t}}.&\sum\limits_{i = 1}^K {{{ c}_i}}  \le {{ \bar{R} }_{k,{\mathrm{c}}}},\forall k \in \mathcal{K}, \label{constraint_ini_1}\\
& {{ c}_k} \ge 0,\forall k \in \mathcal{K},\label{constraint_ini_2}\\
&\sum\limits_{i = 0}^K {{A_i}} {\mathbf{p}}_i^T{{\mathbf{e}}_n} \le \min \left\{ {b - {I_{\mathrm{L}}},} \right.
\left. {{I_{\mathrm{H}}} - b} \right\},\forall n \in \mathcal{N} ,\label{constraint_ini_3}\\
&\sum\limits_{k = 0}^K {{{\left\| {{{\mathbf{p}}_k}} \right\|}^2}} {\varepsilon _k} \le {P_t},\label{constraint_ini_4}\\
&{{\mathbf{h}}_k} = {{{\mathbf{\hat h}}}_k}{\mathrm{ + }}\Delta {{\mathbf{h}}_k},\Delta {{\mathbf{h}}_k} \in {{\mathcal{H}}_{k}},\forall k \in {\mathcal{K}}.\label{constraint_ini_5}
\end{align}
\end{subequations}
In problem \eqref{problem_ini}, $P_t$ represents the total transmit power.
Constraint \eqref{constraint_ini_1} and constraint \eqref{constraint_ini_2} guarantee the decodability and non-negative rate of the common stream at each user, respectively. 
The optical power constraint and the electric power constraint are respectively represented in constraints \eqref{constraint_ini_3} and  \eqref{constraint_ini_4}.
Overall, problem \eqref{problem_ini} is a non-convex optimization problem due to the intricate rate expression. 
It is highly challenging to directly resolve problem \eqref{problem_ini}.
To resolve the problem \eqref{problem_ini}, in the next section, we present a iterative CCCP-based algorithm.
\begin{figure*}[ht]
      \centering
	\includegraphics[width=13.11cm]{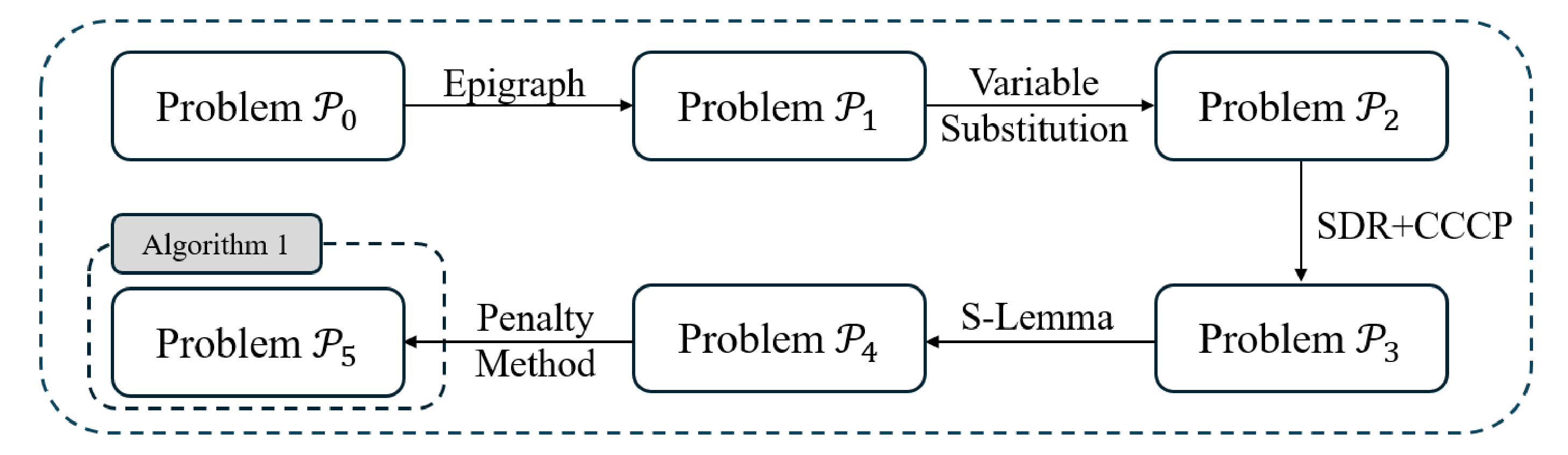}
\caption{Proposed optimization framework to solve $\mathcal{P}_0$.}
   \label{al_framework}
\end{figure*}
\section{BEAMFORMING DESIGN WITH IMPERFECT CSI}
In this section, inspired by the optimization algorithm proposed in \cite{2020_zhou_irs_sdr_cccp}, we address the non-convexities in problem $\mathcal{P}_0$ and transform it into a convex subproblem.
Specifically, we first apply SDR and CCCP algorithms to deal with the non-convexity in the problem $\mathcal{P}_0$. 
Then, we transform the semi-infinite constraints imposed by the imperfect CSIT into finite ones, and we introduce a penalty method to address the rank-one constraint introduced by SDR. 
Finally, the transformed convex subproblem is solved iteratively to obtain a near optimal solution. The  proposed optimization framework is illustrated in Fig. \ref{al_framework}. 
The detailed derivation process is described in the following subsections.

\subsection{ Semidefinite Relaxation }
In MMF problem \eqref{problem_ini}, since the objective function is a minimum function, we cannot directly solve it.
Here, we first introduce an auxiliary variable $t$ to transform problem \eqref{problem_ini} equivalently to :
\begin{subequations}\label{problem_transform1}
\begin{align} 
\mathbf{\mathcal{P}_1 :}   \ \ \ \mathop {\max }\limits_{\left\{ {{\mathbf{p}}_i} \right\}_{i = 0}^K,\left\{ c_k \right\}_{k = 1}^K,t} \ \ \ & t \\
{\mathrm{s}}.{\mathrm{t}}. \ \ \ & c_k + \bar{R} _{k,{\mathrm{p}}} \ge t,\forall k \in {\mathcal{K}},\label{constraint_t1_1}\\
& \eqref{constraint_ini_1}-\eqref{constraint_ini_5} . \nonumber
\end{align}
\end{subequations}
It is noteworthy that constraint \eqref{constraint_t1_1} is equivalent to $\min_{k\in \mathcal{K} } \left \{ c_k +  \bar{R} _{k,{\mathrm{p}}} \right \}\ge t $, and the inequation must hold with equality at optimum.
Thus, the equivalence between \eqref{problem_ini} and \eqref{problem_transform1} is guaranteed.

The non-convexity of the constraints \eqref{constraint_ini_1} and \eqref{constraint_t1_1} makes problem \eqref{problem_transform1} computationally intractable. 
In order to address this issue, several auxiliary variables are firstly introduced as below
\begin{subequations}\label{introduce_variable}
\begin{align}
&\widehat {\mathbf{p}}    =  \left [ \mathbf{p}_0^T,\dots ,\mathbf{p}_K^T\right ]  ^T ,\\
& \widehat {\mathbf{d}}  = \left [ \varepsilon_0^{1/2},\dots,\varepsilon_K^{1/2} \right ] ^T \otimes \mathbf{1}_N,\\
&{{\mathbf{a}}_n}  = {\left[ {{A_0}, \ldots, {A_K}} \right]^T} \otimes {{\mathbf{e}}_n},\\
& \Delta \overline{\mathbf{H}}_k   = \tau_0 \Delta\mathbf{h}_k \Delta\mathbf{h}_k^T ,\\
& \Delta \underline{\mathbf{H}}_k   = 2 \pi \varepsilon_0 \Delta\mathbf{h}_k \Delta\mathbf{h}_k^T ,\\
&{{\mathbf{H}}_{{\mathrm{c}},k}}   = {\mathrm{diag}}\left\{ {{\tau _0}, \ldots ,{\tau _K}} \right\} \otimes \left( {{{\mathbf{h}}_k}{\mathbf{h}}_k^T} \right),\\
&{\overline {\mathbf{H}} _{{\mathrm{c}},k}}{\mathrm{ }}  = 2\pi {\mathrm{diag}}\left\{ {0,{\varepsilon _1}, \ldots ,{\varepsilon _K}} \right\} \otimes \left( {{{\mathbf{h}}_k}{\mathbf{h}}_k^T} \right),\\
&{\widehat {\mathbf{H}}_{{\mathrm{c}},k}}   = {\mathrm{diag}}\left\{ {0,{\tau _1}, \ldots ,{\tau _K}} \right\} \otimes \left( {{{\mathbf{h}}_k}{\mathbf{h}}_k^T} \right),\\
&{\widehat {\mathbf{H}}_{{\mathrm{p}},k}}  = 2\pi {\mathrm{diag}}\left\{ {0,{\varepsilon _1}, \ldots, {\varepsilon _{i - 1}},0,{\varepsilon _{i + 1}}, \ldots ,{\varepsilon _K}} \right\}\nonumber\\
 & \qquad~~~ \otimes \left( {{{\mathbf{h}}_k}{\mathbf{h}}_k^T} \right),
\end{align}
\end{subequations}
where $\otimes$ denotes Kronecker-product. \par
Based on the definitions in \eqref{introduce_variable} and the lower bound of the achievable rates derived in \eqref{common_rate_lower_bound} and \eqref{private_rate_lower_bound}, problem \eqref{problem_transform1} can be equivalently reformulated as
\begin{subequations}\label{problem_t2}
\begin{align}
\mathbf{\mathcal{P}_2 :}   \ \ \ & \mathop {\max }\limits_{ \hat{\mathbf{p}},\left\{ {{c_k}} \right\}_{k = 1}^K,t}\ \ \   t\\
{\mathrm{s}}.{\mathrm{t}}.\ \ \ &\frac{1}{2}{\log _2}\left( {2\pi \sigma _k^2 + {{\widehat {\mathbf{p}}}^T}{{\mathbf{H}}_{{\mathrm{c}},k}}\widehat {\mathbf{p}}} \right)\nonumber\\
&- \frac{1}{2}{\log _2}\left( {2\pi \sigma _k^2 + {{\widehat {\mathbf{p}}}^T}{{\overline {\mathbf{H}} }_{{\mathrm{c}},k}}\widehat {\mathbf{p}}} \right) \ge \sum\limits_{i = 1}^K c_i ,\forall k \in {\mathcal{K}},\label{constraint_t2_1}\\
&\frac{1}{2}{\log _2}\left( {2\pi \sigma _k^2 + {{\widehat {\mathbf{p}}}^T}{{\widehat {\mathbf{H}}}_{{\mathrm{c}},k}}\widehat {\mathbf{p}}} + \mathbf{p}_0^T \Delta \overline{\mathbf{H}}_k \mathbf{p}_0\right)\nonumber\\
&- \frac{1}{2}{\log _2}\left( {2\pi \sigma _k^2 + {{\widehat {\mathbf{p}}}^T}{{\widehat {\mathbf{H}}}_{{\mathrm{p}},k}}\widehat {\mathbf{p}}}  + \mathbf{p}_0^T \Delta \underline{\mathbf{H}}_k \mathbf{p}_0\right) \nonumber\\
& \geq t - {c_k},\forall k \in {\mathcal{K}},\label{constraint_t2_2}\\
&{\widehat {\mathbf{p}}^T}{{\mathbf{a}}_n}{\mathbf{a}}_n^T\widehat {\mathbf{p}} \le \min \left\{ {{{\left( {b - {I_{\mathrm{L}}}} \right)}^2},{{\left( {{I_{\mathrm{H}}} - b} \right)}^2}} \right\},\nonumber\\
& \forall n \in {\mathcal{N}},\label{constraint_t2_3}\\
&{\left\| {{\mathbf{\widehat p}} \odot \left( {\widehat {\mathbf{d}}{{\widehat {\mathbf{d}}}^T}} \right)} \right\|^2} \le {P_t},\label{constraint_t2_4}\\
& \eqref{constraint_ini_2}, \eqref{constraint_ini_5} ,\nonumber 
\end{align}
\end{subequations}
where $\odot$ denotes Hadamard product. 

Constraints \eqref{constraint_t2_1} and \eqref{constraint_t2_2} remain non-convex.
Observing the quadratic form in these two constraints, it is therefore appropriate to apply the SDR method to relax the constraints. 
Specifically, by defining $\widehat {\mathbf{P}}  =  \widehat {\mathbf{p}}{\widehat {\mathbf{p}}^T} $, constraint \eqref{constraint_t2_1} can be equivalently rewritten as
\begin{align}\label{constraint_t1_sdr_renewed}
&\underbrace {\frac{1}{2}{{\log }_2}\left( {2\pi \sigma _k^2 + {\mathrm{Tr}}(\widehat {\mathbf{P}}{{\mathbf{H}}_{{\mathrm{c}},k}})} 
 \right)}_{{\mathrm{concave}}}\nonumber\\
&  \underbrace {-\frac{1}{2}{{\log }_2}\left( {2\pi \sigma _k^2 + {\mathrm{Tr}}(\widehat {\mathbf{P}}{{\overline {\mathbf{H}} }_{{\mathrm{c}},k}})} \right)}_{{\mathrm{convex}}} \ge \sum\limits_{i = 1}^K {c_i}.
\end{align}
Similarly, by defining $\mathbf{P}_0 = \mathbf{p}_0\mathbf{p}_0^T$, constraint \eqref{constraint_t2_2} can be reformulated as follows
\begin{align}\label{constraint_t2_sdr_renewed}
& \underbrace{\frac{1}{2}\log_{2}{\left ( 2\pi \sigma _k^2 +\mathrm{Tr}(\widehat {\mathbf{P}}{{\widehat {\mathbf{H}}}_{{\mathrm{c}},k}}) +  \mathrm{Tr}(\mathbf{P}_0 \Delta \overline{\mathbf{H}}_k  ) \right ) }  }_{\mathrm{concave} } \nonumber \\
&   \underbrace{- \frac{1}{2}\log_{2}{\left ( 2\pi \sigma _k^2 +\mathrm{Tr}(\widehat {\mathbf{P}}{{\widehat {\mathbf{H}}}_{{\mathrm{p}},k}}) +  \mathrm{Tr}(\mathbf{P}_0 \Delta \underline{\mathbf{H}}_k  ) ) \right ) }  }_{\mathrm{convex} } \geq t-c_k . 
\end{align}
Constraints \eqref{constraint_t1_sdr_renewed} and \eqref{constraint_t2_sdr_renewed} remain non-convex, but the left-hand side of both inequalities comprises both concave and convex terms.
To deal with this non-convexity, we further define ${\mathbf{P}}_i = {{\mathbf{p}}_i}{\mathbf{p}}_i^T$, ${\mathbf{\Phi }}  = \sum\nolimits_{i = 0}^K {{\tau _i}{{\mathbf{P}}_i}} $, ${\mathbf{\bar \Phi }}   = \sum\nolimits_{j = 1}^K {2\pi {\varepsilon _j}{{\mathbf{P}}_j}}  $ and ${{\mathbf{H}}_k}   = {{\mathbf{h}}_k}{\mathbf{h}}_k^T$, and introduce the following inequalities for \eqref{constraint_t1_sdr_renewed}
\begin{subequations}\label{ro_common_rate_variable}
\begin{align}
&\exp \left( {{x_{k,{\mathrm{c}}}}} \right) \le {\mathrm{Tr}}\left( {{\mathbf{\Phi }}{{\mathbf{H}}_k}} \right) + 2\pi \sigma _k^2,\label{common_numerator_variable}\\
&\exp \left( {{y_{k,{\mathrm{c}}}}} \right) \ge {\mathrm{Tr}}\left( {{\mathbf{\bar \Phi }}{{\mathbf{H}}_k}} \right) + 2\pi \sigma _k^2,\label{common_denominator_variable}
\end{align}
\end{subequations}
where $x_{k,{\mathrm{c}}}$ and $y_{k,{\mathrm{c}}}$ are introduced slack variables.
Based on \eqref{ro_common_rate_variable}, constraint \eqref{constraint_t1_sdr_renewed} can be approximated as 
\begin{align}\label{constraint_t1_final_version}
\sum\limits_{i = 1}^K {{c_i}}  - \frac{1}{2}{\log _2}\exp \left( {{x_{k,{\mathrm{c}}}} - {y_{k,{\mathrm{c}}}}} \right) \le 0,\forall k \in {\mathcal{K}}.
\end{align}
Note that the transformation from constraint \eqref{constraint_t2_1} to constraint \eqref{constraint_t1_final_version} is a feasible relaxation.

Following the same procedure, we further define ${\mathbf{Q}}   = \sum\nolimits_{i = 1}^K {{\tau _i}{{\mathbf{P}}_i}} $, ${{{\mathbf{\tilde P}}}_0}  =  {\tau _0}{{\mathbf{P}}_0}$,
${\mathbf{\bar Q}}  = \sum\nolimits_{j = 1,j \ne k}^K {2\pi {\varepsilon _j}{{\mathbf{P}}_j}}  $,  ${{{\mathbf{\bar P}}}_0}  = 2\pi {\varepsilon _0}{{\mathbf{P}}_0}$ and $\Delta {{\mathbf{H}}_k}   = \Delta {{\mathbf{h}}_k}\Delta {\mathbf{h}}_k^T$.
For constraint \eqref{constraint_t2_sdr_renewed}, we also introduce the following inequalities:
\begin{subequations}\label{ro_private_rate_variable}
\begin{align}
\exp \left( {{x_{k,{\mathrm{p}}}}} \right) &\le {\mathrm{Tr}}\left( {{\mathbf{Q}}{{\mathbf{H}}_k}} \right) + {\mathrm{Tr}}\left( {{{{\mathbf{\tilde P}}}_0}\Delta {{\mathbf{H}}_k}} \right)
 + 2\pi \sigma _k^2,\label{private_fenzi_variable}\\
\exp \left( {{y_{k,{\mathrm{p}}}}} \right) &\ge {\mathrm{Tr}}\left( {{\mathbf{\bar Q}}{{\mathbf{H}}_k}} \right) + {\mathrm{Tr}}\left( {{{{\mathbf{\bar P}}}_0}\Delta {{\mathbf{H}}_k}} \right) + 2\pi \sigma _k^2,\label{private_fenmu_variable}
\end{align}
\end{subequations}
where $x_{k,{\mathrm{p}}}$ and $y_{k,{\mathrm{p}}}$ are slack variables.
Constraint \eqref{constraint_t2_sdr_renewed} is then reformulated as 
\begin{align}\label{constraint_t2_final_version}
t - {{c}_k} - \frac{1}{2}{\log _2}\exp \left( {{x_{k,{\mathrm{p}}}} - {y_{k,{\mathrm{p}}}}} \right) \le 0,\forall k \in {\mathcal{K}}.
\end{align}

By further introducing slack variables $\left( {{p_{k,{\mathrm{c}}}},{q_{k,{\mathrm{c}}}},{p_{k,{\mathrm{p}}}},{q_{k,{\mathrm{p}}}}} \right)$ for \eqref{ro_common_rate_variable} and \eqref{ro_private_rate_variable}, problem \eqref{problem_t2} is then equivalently reformulated as 
\begin{subequations}\label{problem_t3}
\begin{align}
\mathbf{\mathcal{P}_3 :}   \ \ & \ \mathop {\max }\limits_{\mathbf{\Omega },t}\ \ \     t   \\
{\mathrm{s}}{\mathrm{.t}}{\mathrm{.}}\ \ \  &
\sum\limits_{i = 1}^K {{c_i}}  - \frac{1}{2}{\log _2}\exp \left( {{x_{k,{\mathrm{c}}}} - {y_{k,{\mathrm{c}}}}} \right) \le 0,\forall k \in {\mathcal{K}}, \label{constraint1_after_slack}\\
& t - {{c}_k} - \frac{1}{2}{\log _2}\exp \left( {{x_{k,{\mathrm{p}}}} - {y_{k,{\mathrm{p}}}}} \right) \le 0,\forall k \in {\mathcal{K}},\label{constraint2_after_slack}\\
&\exp \left( {{x_{k,{\mathrm{c}}}}} \right) \le {p_{k,{\mathrm{c}}}},\exp \left( {{x_{k,{\mathrm{p}}}}} \right) \le {p_{k,{\mathrm{p}}}},\forall k \in {\mathcal{K}},\label{constraint_final_2}\\
&\exp \left( {{y_{k,{\mathrm{c}}}}} \right) \ge {q_{k,{\mathrm{c}}}},\exp \left( {{y_{k,{\mathrm{p}}}}} \right) \ge {q_{k,{\mathrm{p}}}},\forall k \in {\mathcal{K}},\label{constraint_final_3}\\
&{\mathrm{Tr}}\left( {{\mathbf{\Phi }}{{\mathbf{H}}_k}} \right) + 2\pi \sigma _k^2 \ge {p_{k,{\mathrm{c}}}},\forall k \in {\mathcal{K}},\label{constraint_final_4}\\
&{\mathrm{Tr}}\left( {{\mathbf{\bar \Phi }}{{\mathbf{H}}_k}} \right) + 2\pi \sigma _k^2 \le {q_{k,{\mathrm{c}}}},\forall k \in {\mathcal{K}},\label{constraint_final_5}\\
&{\mathrm{Tr}}\left( {{\mathbf{Q}}{{\mathbf{H}}_k}} \right) + {\mathrm{Tr}}\left( {{{{\mathbf{\tilde P}}}_0}\Delta {{\mathbf{H}}_k}} \right) + 2\pi \sigma _k^2 \nonumber\\
&\ge {p_{k,{\mathrm{p}}}},\forall k \in {\mathcal{K}},\label{constraint_final_6}\\
&{\mathrm{Tr}}\left( {{\mathbf{\bar Q}}{{\mathbf{H}}_k}} \right) + {\mathrm{Tr}}\left( {{{{\mathbf{\bar P}}}_0}\Delta {{\mathbf{H}}_k}} \right) + 2\pi \sigma _k^2\nonumber\\ 
&\le {q_{k,{\mathrm{p}}}},\forall k \in {\mathcal{K}},\label{constraint_final_7}\\
&{x_{k,{\mathrm{c}}}} \ge {y_{k,{\mathrm{c}}}},{x_{k,{\mathrm{p}}}} \ge {y_{k,{\mathrm{p}}}},\forall k \in {\mathcal{K}},\label{constraint_final_8}\\
&\sum\limits_{i = 0}^K {{\varepsilon _i}} {\mathrm{Tr}}\left( {{\mathbf{P}}_i^{}} \right) \le {P_t} ,\label{power_constraint_1}\\ 
&\sum\limits_{i = 0}^K {A_i^2{\mathrm{Tr}}\left( {{{\mathbf{P}}_i}{{\mathbf{e}}_n}{{\mathbf{e}}_n}^T} \right)}  \le min\left\{ {{{\left( {b - {I_{\mathrm{L}}}} \right)}^2},} \right.\nonumber\\
&\left. {{{\left( {{I_{\mathrm{H}}} - b} \right)}^2}} \right\},\forall n \in \mathcal{N},\label{power_constraint_2}\\
&{{\mathbf{P}}_i} \succeq {\mathbf{0}},\forall i \in {\mathcal{I}},\label{constraint_final_11} \\
& \mathrm{Rank}\left ( \mathbf{P}_i  \right )   = 1, \ \forall  i \in \mathcal{I} ,\label{constraint_final_12}\\
& \eqref{constraint_ini_2}, \eqref{constraint_ini_5} ,\nonumber 
\end{align}
\end{subequations}
where $\mathbf{\Omega} = \left\{ {{{\mathbf{P}}_i},{c_k},{x_{k,{\mathrm{c}}}},{y_{k,{\mathrm{c}}}},{x_{k,{\mathrm{p}}}},{y_{k,{\mathrm{p}}}},{p_{k,{\mathrm{c}}}},} \right.$ $\left. {{q_{k,{\mathrm{c}}}},{p_{k,{\mathrm{p}}}},{q_{k,{\mathrm{p}}}}\left| \; {\forall i \in \mathcal{I},\forall k \in \mathcal{K}} \right.} \right\}$ represents the set of variables.
Constraint \eqref{constraint_final_8} ensures that the lower bounds of the achievable rates at each user are guaranteed to be non-negative.
Constraints \eqref{constraint_final_11} and \eqref{constraint_final_12} ensure that ${\mathbf{P}}_i$ is both positive semi-definite and rank-one.

\subsection{Transformation of Semi-Infinite Constraints} 
Constraints \eqref{constraint_final_4} - \eqref{constraint_final_7} involve $\mathbf{H}_k =  \mathbf{h}_k \mathbf{h}_k^T$ and $\Delta\mathbf{H}_k  = \Delta\mathbf{h}_k \Delta\mathbf{h}_k^T $, $\forall k \in \mathcal{K}$. 
By substituting \eqref{constraint_ini_5} into \eqref{constraint_final_4} - \eqref{constraint_final_7}, respectively, we have
\begin{subequations}\label{constraint_before_s_lemma}
\begin{align}
\Delta {\mathbf{h}}_k^T{\mathbf{\Phi }}\Delta {{\mathbf{h}}_k}  + & 2\Delta {\mathbf{h}}_k^T{\mathbf{\Phi }}{{{\mathbf{\hat h}}}_k} + {\mathbf{\hat h}}_k^T{\mathbf{\Phi }}{{{\mathbf{\hat h}}}_k} \nonumber\\
 & + 2\pi \sigma _k^2 - {p_{k,{\mathrm{c}}}} \ge 0, \forall k \in \mathcal{K},\\
\Delta {\mathbf{h}}_k^T{\mathbf{\bar \Phi }}\Delta {{\mathbf{h}}_k} + & 2\Delta {\mathbf{h}}_k^T{\mathbf{\bar \Phi }}{{{\mathbf{\hat h}}}_k} + {\mathbf{\hat h}}_k^T{\mathbf{\bar \Phi }}{{{\mathbf{\hat h}}}_k}\nonumber\\ 
 & + 2\pi \sigma _k^2 - {q_{k,{\mathrm{c}}}} \le 0, \forall k \in \mathcal{K},\\
\Delta {\mathbf{h}}_k^T{\mathbf{R}}\Delta {{\mathbf{h}}_k} + & 2\Delta {\mathbf{h}}_k^T{\mathbf{Q}}{{{\mathbf{\hat h}}}_k} + {\mathbf{\hat h}}_k^T{\mathbf{Q}}{{{\mathbf{\hat h}}}_k}\nonumber\\ 
 & + 2\pi \sigma _k^2 - {p_{k,{\mathrm{p}}}} \ge 0, \forall k \in \mathcal{K},\\
\Delta {\mathbf{h}}_k^T{\mathbf{\bar R}}\Delta {{\mathbf{h}}_k} + & 2\Delta {\mathbf{h}}_k^T{\mathbf{\bar Q}}{{{\mathbf{\hat h}}}_k} + {\mathbf{\hat h}}_k^T{\mathbf{\bar Q}}{{{\mathbf{\hat h}}}_k}\nonumber\\
 & + 2\pi \sigma _k^2 - {q_{k,{\mathrm{p}}}} \le 0, \forall k \in \mathcal{K}, \\
 &  \Delta {{\mathbf{h}}_k} \in {{\mathcal{H}}_{k}},\forall k \in {\mathcal{K}},
\end{align}
\end{subequations}
where ${\mathbf{R}}  = {{{\mathbf{\tilde P}}}_0} + {\mathbf{Q}}$ and ${\mathbf{\bar R}}   = {{{\mathbf{\bar P}}}_0} + {\mathbf{\bar Q}}$.\par
Due to the CSI estimation errors $\Delta {{\mathbf{h}}_k} \in {{\mathcal{H}}_k}$, constraint \eqref{constraint_before_s_lemma} involves infinite constraints, which is intractable. 
In optimization theory, such problem is known as semi-infinite problem (SIP) or robust optimization. 
To deal with this issue, we transform \eqref{constraint_before_s_lemma} into finite linear matrix inequality (LMI) constraints by introducing the following lemma.

\textbf{Lemma 1 ($\mathcal{S}$-lemma) }\cite{S_Boyd} \textbf{:}
Define a function ${f_m}\left( {\mathbf{z}} \right)$, $m = \left\{ {1,2} \right\}$ as follows : 
\begin{align}
{f_m}\left( {\mathbf{z}} \right) = {{\mathbf{z}}^T}{{\mathbf{A}}_m}{\mathbf{z}} + 2{\mathrm{Re}}\left\{ {{\mathbf{r}}_m^T{\mathbf{z}}} \right\} + {z_m},
\end{align}
where ${\mathbf{z}} \in {{\mathbb{R}}^{N \times 1}},{{\mathbf{A}}_m}= {\mathbf{A}}_m^T \in {{\mathbb{R}}^{N \times N}},{{\mathbf{r}}_m} \in {{\mathbb{R}}^{N \times 1}}$, and ${z_m} \in {\mathbb{R}}$.
Then, the implication ${f_1}\left( {\mathbf{z}} \right) \le 0 \Rightarrow {f_2}\left( {\mathbf{z}} \right) \le 0$ holds if and only if there exits a variable $\lambda  \ge 0$ such that
\begin{align}
\lambda \left[ {\begin{array}{*{20}{c}}
{{{\mathbf{A}}_1}}&{{{\mathbf{r}}_1}}\\
{{\mathbf{r}}_1^T}&{{z_1}}
\end{array}} \right] - \left[ {\begin{array}{*{20}{c}}
{{{\mathbf{A}}_2}}&{{{\mathbf{r}}_2}}\\
{{\mathbf{r}}_2^T}&{{z_2}}
\end{array}} \right] \succeq {\mathbf{0}}.
\end{align}
Such an equivalence is guaranteed if there exists a point $\widehat {\mathbf{z}}$ satisfying ${f_m}\left( {\widehat {\mathbf{z}}} \right) \le 0$, $m = \left\{ {1,2} \right\}$.

By applying $\mathcal{S}$-lemma to constraint \eqref{constraint_before_s_lemma}, the following convex LMI constraints can be derived conservatively :
\begin{subequations}\label{constraint_after_s_lemma}
\begin{align}
&\left[ {\begin{array}{*{20}{c}}
{{u_{k,{\mathrm{c}}}}{{\mathbf{I}}_N} + {\mathbf{\Phi }}}&{{\mathbf{\Phi }}{{{\mathbf{\hat h}}}_k}}\\
{{\mathbf{\hat h}}_k^T{\mathbf{\Phi }}}&{ - {u_{k,{\mathrm{c}}}}{\upsilon _k} + {\mathbf{\hat h}}_k^T{\mathbf{\Phi }}{{{\mathbf{\hat h}}}_k} + 2\pi \sigma _k^2 - {p_{k,{\mathrm{c}}}}}
\end{array}} \right]\succeq{\mathbf{0}},\\
&\left[ {\begin{array}{*{20}{c}}
{{\lambda _{k,{\mathrm{c}}}}{{\mathbf{I}}_N} - {\mathbf{\bar \Phi }}}&{ - {\mathbf{\bar \Phi }}{{{\mathbf{\hat h}}}_k}}\\
{ - {\mathbf{\hat h}}_k^T{\mathbf{\bar \Phi }}}&{ - {\lambda _{k,{\mathrm{c}}}}{\upsilon _k} - {\mathbf{\hat h}}_k^T{\mathbf{\bar \Phi }}{{{\mathbf{\hat h}}}_k} - 2\pi \sigma _k^2 + {q_{k,{\mathrm{c}}}}}
\end{array}} \right]\succeq {\mathbf{0}},\\
&\left[ {\begin{array}{*{20}{c}}
{{u_{k,{\mathrm{p}}}}{{\mathbf{I}}_N} + {\mathbf{R}}}&{{\mathbf{Q}}{{{\mathbf{\hat h}}}_k}}\\
{{\mathbf{\hat h}}_k^T{\mathbf{Q}}}&{ - {u_{k,{\mathrm{p}}}}{\upsilon _k} + {\mathbf{\hat h}}_k^T{\mathbf{Q}}{{{\mathbf{\hat h}}}_k} + 2\pi \sigma _k^2 - {p_{k,{\mathrm{p}}}}}
\end{array}} \right]\succeq {\mathbf{0}},\\
&\left[ {\begin{array}{*{20}{c}}
{{\lambda _{k,{\mathrm{p}}}}{{\mathbf{I}}_N} - {\mathbf{\bar R}}}&{ - {\mathbf{\bar Q}}{{{\mathbf{\hat h}}}_k}}\\
{ - {\mathbf{\hat h}}_k^T{\mathbf{\bar Q}}}&{ - {\lambda _{k,{\mathrm{p}}}}{\upsilon _k} - {\mathbf{\hat h}}_k^T{\mathbf{\bar Q}}{{{\mathbf{\hat h}}}_k} - 2\pi \sigma _k^2 + {q_{k,{\mathrm{p}}}}}
\end{array}} \right]\succeq {\mathbf{0}},
\end{align}
\end{subequations}
where ${u_{k,{\mathrm{c}}}} \ge 0, {{\lambda}_{k,{\mathrm{c}}}} \ge 0, {u_{k,{\mathrm{p}}}} \ge 0,{{\lambda}_{k,{\mathrm{p}}}} \ge 0$ are auxiliary variables. 
It is obvious that constraint \eqref{constraint_after_s_lemma} only involves a finite number of convex constraints.
Thus, with the help of $\mathcal{S}$-lemma, we successfully transform the semi-infinite constraints in \eqref{constraint_before_s_lemma} into finite constraints in \eqref{constraint_after_s_lemma}.

\subsection{CCCP-Based Transformation and Overall Algorithm} 
After replacing \eqref{constraint_final_4}--\eqref{constraint_final_7} with \eqref{constraint_after_s_lemma}, only constraint \eqref{constraint_final_3} is non-convex in problem \eqref{problem_t3}.
In this subsection, an iterative CCCP-based transformation is proposed to solve this non-convexity.

To do so, we linearize the left-hand side of constraint \eqref{constraint_final_3} by Taylor's first-order expansion.
For a feasible point of $y_{k,c}^{ \left [ n-1 \right ] }$ and $y_{k,p}^{ \left [ n-1 \right ] }$ at the $(n-1)$th iteration, \eqref{constraint_final_3} is approximated as
\begin{subequations}\label{tylor}
\begin{align}
&e^{ {y_{k,{\mathrm{c}}}^{\left[ n-1 \right]}} }\left( {1 + {y_{k,{\mathrm{c}}}} - y_{k,{\mathrm{c}}}^{\left[ n-1 \right]}} \right) \ge {q_{k,{\mathrm{c}}}},\\
&e^{ {y_{k,{\mathrm{p}}}^{\left[ n-1 \right]}} }\left( {1 + {y_{k,{\mathrm{p}}}} - y_{k,{\mathrm{p}}}^{\left[ n-1 \right]}} \right) \ge {q_{k,{\mathrm{p}}}}.
\end{align}
\end{subequations}

Based on \eqref{constraint_after_s_lemma} and \eqref{tylor}, at each iteration, problem \eqref{problem_t3} with non-convexity constraints \eqref{constraint_final_3} and infinite constraints \eqref{constraint_final_4} - \eqref{constraint_final_7} can be strictly approximated to the following subproblem 
\begin{align}\label{problem_final}
\mathbf{\mathcal{P}_4 :}   \ \ \mathop {\max }\limits_{{\mathbf{\Omega }},{\mathbf{\Psi }},t}  \ \ \  &\ t \\
{\mathrm{s}}.{\mathrm{t}}.\ \ \  &\eqref{constraint1_after_slack}, \eqref{constraint2_after_slack},\eqref{constraint_final_2},\eqref{constraint_final_8},\eqref{power_constraint_1},\eqref{power_constraint_2},\nonumber\\
&\eqref{constraint_ini_2},\eqref{constraint_ini_5}, \eqref{constraint_final_11},\eqref{constraint_final_12},\eqref{constraint_after_s_lemma},\eqref{tylor},\nonumber
\end{align}
where ${\mathbf{\Psi }}   = \left\{ {{u_{k,{\mathrm{c}}}},{\lambda _{k,{\mathrm{c}}}},{u_{k,{\mathrm{p}}}},{\lambda _{k,{\mathrm{p}}}}\left| {\forall k \in {\mathcal{K}}} \right.} \right\}$ represents the set of the auxiliary variables used in \eqref{constraint_after_s_lemma}. 

In problem \eqref{problem_final}, if the obtained solution $\mathbf{P}_i^*$, $i \in \mathcal{I}$ is already rank-one, the robust beamforming vectors of problem \eqref{problem_final} can be directly acquired by eigenvalue decomposition (EVD). 
We can also leverage the Gaussian randomization procedure \cite{Luo_2010} to recover a high-quality rank-one solution. 
However, if solution ${\mathbf{P}}_i^*$, $i \in \mathcal{I}$ of \eqref{problem_final} is not rank-one, the solution recaptured by the Gaussian randomization procedure may not satisfy other constraints in the original problem, which may cause further problems. 
For example, we need to further solve a new optimization problem to find the nearest feasible solution, which also increases computational complexity. 
The approximated solution obtained by Gaussian randomization also introduces additional complexity.
To avoid these complexities, in the next subsection, we further apply a penalty method to address the rank-one constraint.

\subsection{A penalty method}

In this subsection, to efficiently address the rank-one constraint \eqref{constraint_final_12}, here we present a penalty-based method. 
We could obtain that constraint \eqref{constraint_final_12} is mathematically equivalent to a difference-of-convex (DC) function-based constraint based on the following proposition 1.

\textbf{Proposition 1 :}
For any positive semidefinite matrix $\mathbf{Z} \in \mathbb{R} ^{N\times N}$, we could obtain that
\begin{align}\label{rankone_constraint}
\mathrm{Rank}(\mathbf{Z}) = 1 \Longleftrightarrow \mathrm{Tr}(\mathbf{Z}) - \left \| \mathbf{Z} \right \| _2 = 0
\end{align}
where $\mathrm{Tr}(\mathbf{Z})  = \sum_{i=1}^{N} \sigma _i(\mathbf{Z})$, $ \left \| \mathbf{Z} \right \| _2 = \sigma _1(\mathbf{Z})$, and $\sigma _i(\mathbf{Z})$ represents the $i$-th largest singular value of the matrix $\mathbf{Z}$. 
\par
By applying proposition 1 to constraint \eqref{constraint_final_12} and augmenting the objective function with a penalty term $F_{\mathrm{penalty} } = \rho\sum_{i=0}^{K} (\mathrm{Tr}(\mathbf{P}_i) - \left \| \mathbf{P}_i \right \| _2) $,
problem \eqref{problem_final} can be transformed to the following form
\begin{align}\label{problem_with_penalty}
 \mathop {\max }\limits_{{\mathbf{\Omega }},{\mathbf{\Psi }},t}  &\ t   + F_{\mathrm{penalty} }\\
{\mathrm{s}}.{\mathrm{t}}.\ \ \ &\eqref{constraint1_after_slack}, \eqref{constraint2_after_slack},\eqref{constraint_final_2},\eqref{constraint_final_8},\eqref{power_constraint_1},\eqref{power_constraint_2},\nonumber\\
&\eqref{constraint_ini_2},\eqref{constraint_ini_5}, \eqref{constraint_final_11},\eqref{constraint_after_s_lemma},\eqref{tylor},\nonumber
\end{align}
where $\rho$ represents a negative penalty parameter which penalizes the objective function if the optimal solution $\mathbf{P}^*_i,\ \forall i \in \mathcal{I}$ is not rank-one. 
It is worth noting that, when the nonnegative component  $\sum_{i=0}^{K} (\mathrm{Tr}(\mathbf{P}_i) - \left \| \mathbf{P}_i \right \| _2)$ in the objective function is zero,  we can obtain an exact rank-one solution $\mathbf{P}^*_i,\ \forall i \in \mathcal{I}$. 
Thus, problem \eqref{problem_final} and problem \eqref{problem_with_penalty} are equivalent \cite{penalty_equal}.
However, with the addition of the penalty term, the objective function \eqref{problem_with_penalty} is no longer convex.
We further employ SCA to address the non-convexity. 
Specifically, by utilizing the first-order Taylor expansion, we obtain a convex upper bound for the penalty term $\left \| \mathbf{P}_i \right \| _2$ as 
\begin{align}
& \left \| \mathbf{P}_i \right \|  _2 \nonumber \\
\approx & \left \| \mathbf{P}_i \right \| ^{[n-1]}_2 + \mathrm{Tr} \left [  \left ( \mathbf{u}^{[n-1]}_{i,max} \right ) ^H\left ( \mathbf{P}_i -\mathbf{P}_i ^{[n-1]} \right ) \left ( \mathbf{u}^{[n-1]}_{i,max} \right ) \right ]  , 
\end{align}
where $\mathbf{P}_i^{\left [ n-1 \right ] }$ represents the optimal solution obtained at iteration $(n-1)$ and $\mathbf{u}^{\left [ n-1 \right ] }_{i,max}$ denotes the  eigenvectors related to the maximum eigenvalues of $\mathbf{P}_i^{\left [ n-1 \right ] }$.
By removing the constant term in the penalty function, we obtain the final convex penalty function
\begin{align}\label{penalty_convex}
F_{\mathrm{penalty} } = \rho\sum_{i=0}^{K} (\mathrm{Tr}(\mathbf{P}_i) - \left ( \mathbf{u}^{[n-1]}_{i,max} \right ) ^H \mathbf{P}_i \mathbf{u}^{[n-1]}_{i,max} ) .
\end{align}
We can then solve problem \eqref{problem_with_penalty} via a sequence of convex subproblems. 
Specifically, with the optimal solution ${\mathbf{P}_i^*}^{\left [ n-1 \right ] },y_{k,c}^{\left [ n-1 \right ] },y_{k,p}^{\left [ n-1 \right ] }$ obtained at iteration $(n-1)$, we solve the following subproblem at iteration $n$ 
\begin{align}\label{subproblem_with_penalty}
\mathbf{\mathcal{P}_5 :}  \ \ \mathop {\max }\limits_{{\mathbf{\Omega }},{\mathbf{\Psi }},t} \ \ \  &\ t + F_{\mathrm{penalty} } \\
{\mathrm{s}}.{\mathrm{t}}.\ \ \ &\eqref{constraint1_after_slack}, \eqref{constraint2_after_slack},\eqref{constraint_final_2},\eqref{constraint_final_8},\eqref{power_constraint_1},\eqref{power_constraint_2},\nonumber\\
&\eqref{constraint_ini_2},\eqref{constraint_ini_5}, \eqref{constraint_final_11},\eqref{constraint_after_s_lemma},\eqref{tylor},\nonumber
\end{align}
Problem \eqref{subproblem_with_penalty} is convex and we can solve it via classic convex optimization solvers.\par

The overall optimization algorithm to address problem \eqref{problem_ini} is illustrated in Algorithm \ref{algorithm_penalty}.
In each iteration, we solve problem \eqref{subproblem_with_penalty} to update $\mathbf{P}_i^{ \left [ n \right ] }$, $t^{ \left [ n \right ] }$, $y_{k,{\mathrm{c}}}^{ \left [ n \right ] }$ and $y_{k,{\mathrm{p}}}^{ \left [ n \right ] }$. 
The iteration ends until the convergence of the objective function. 
As $P_i^*$ is guaranteed to be rank-one by the penalty term, we can then use EVD to recover the near-optimal beamforming vector $p_i^*$ from $P_i^*$.

It is important to note that adding the penalty function term to the objective function does not increase the algorithm's complexity while guaranteeing a rank-one matrix solution.
It is worth noting that a proper $\rho$ should be chosen based on the specific optimization problem and it varies in different optimization targets in different references \cite{ISAC_ref1,sdr_penalty}.
An appropriate penalty coefficient $\rho$ should be chosen to facilitate the stable and rapid convergence of the algorithm.
Meanwhile, the choice of penalty coefficient $\rho$ should not be too small, otherwise the penalty term would dominate the objective function and its influence to the MMF rate would be non-negligible.
Furthermore, an appropriate penalty coefficient $\rho$ should ensure that the penalty term $F_{\mathrm{penalty}}$ approaches zero closely \cite{2024_kexin}.
Thus, we initialize $\rho$ within a small interval $\left [ -0.1,-0.01 \right ] $ and choose a value which leads to a faster convergence speed, i.e., $\rho = -0.01$.
Then, after several iterations, we gradually increase $\rho$ to ensure stable convergence of the algorithm. 
This process continues until $\rho$ reaches a sufficiently large value, yielding a feasible first-order solution.
\begin{algorithm}[t!]
	\caption{ Penalty Function Method Algorithm }
	\begin{algorithmic}[1]
		\State \textbf{Input}: Initialize $n=0$, feasible $\mathbf{P}_i^{\left[n \right]}$, $i \in \mathcal{I}$, define ${\ t ^{\left[n \right]}} \leftarrow 0$, the penalty factor $\rho$, feasible points $y_{k,{\mathrm{c}}}^{\left[ n \right]}$, $y_{k,{\mathrm{p}}}^{\left[ n \right]}$, $k \in \mathcal{K}$, convergence accuracy $\zeta$;
		\State \textbf{Repeat}
           \State \hspace*{1em} $n \leftarrow n + 1$;
           \State \hspace*{1em} Solve problem \eqref{subproblem_with_penalty} with given $\mathbf{P}_i^{\left[n-1 \right]}$, $i \in \mathcal{I}$, $y_{k,{\mathrm{c}}}^{\left[ n-1\right]}$   
           \Statex \hspace*{1em} and $y_{k,{\mathrm{p}}}^{\left[ n -1\right]}$, $k \in \mathcal{K}$ to obtain the objective function 
           \Statex \hspace*{1em} value as $t^* + F_{\mathrm{penalty}}^*$ and the corresponding solution  
           \Statex \hspace*{1em} as ${\mathbf{P}}_i^*$, $ i \in \mathcal{I}$ , $y_{k,{\mathrm{c}}}^*$ and $y_{k,{\mathrm{p}}}^*$, $ k \in \mathcal{K}$.
          \State \hspace*{1em} Update $ t ^{\left[ n \right]} \leftarrow {\ t ^*}$, $\mathbf{P}_i^{\left[ n \right]} \leftarrow {\mathbf{P}}_i^*$, $i \in \mathcal{I}$, $y_{k,{\mathrm{c}}}^{\left[ n \right]} \leftarrow y_{k,{\mathrm{c}}}^*$, \State \hspace*{1em} $y_{k,{\mathrm{p}}}^{\left[ n \right]}  \leftarrow  y_{k,{\mathrm{p}}}^*$, $k \in \mathcal{K}$;
           
		\State  \textbf{until} $ \left |  \left ( t^{\left [ n+1 \right ] } + F_{\mathrm{penalty}}^{\left [ n+1 \right ] }  \right ) - \left ( t^{\left [ n \right ] } + F_{\mathrm{penalty}}^{\left [ n \right ] } \right )  \right | < \zeta $;
        \State  \textbf{Output}: $\mathbf{p}_i^*= \mathrm{EVD} \left ( \mathbf{P}_i^* \right ) $, ${\ t ^*}$.
	\end{algorithmic} \label{algorithm_penalty}
\end{algorithm}

\subsection{Convergence and Computational Complexity Analysis}

\subsubsection{Convergence Analysis}
Due to the fact that the solution attained by solving problem \eqref{subproblem_with_penalty} at iteration $n$ is also a feasible point of problem \eqref{subproblem_with_penalty} at iteration $(n+1)$, the objective function of \eqref{subproblem_with_penalty} is guaranteed to increase monotonically.
Meanwhile, there exist power constraints \eqref{power_constraint_1} and \eqref{power_constraint_2}, which ensures that the objective function $\left \{ t^{\left [ n \right ] }  + F^{\left [ n \right ] } _{\mathrm{penalty} }\right \} _{n=1}^{n = \infty }$ is bounded above. 
Thus, the convergence of the proposed algorithm is guaranteed.

\subsubsection{Computational Complexity Analysis}
In each iteration, the convex problem \eqref{subproblem_with_penalty} is solved. 
Problem \eqref{subproblem_with_penalty} contains $1+13K+(K+1)N^2$ optimization variables. 
This problem can be resolved by applying the interior point method with computational complexity $\mathcal{O} \left (  \left [ KN^2  \right ]^{3.5} \right ) $.
The total number of iterations needed for the convergence can be approximated as $\mathcal{O} \left (  \log_{}{ \left ( \zeta^{-1} \right ) }  \right ) $ \cite{linear_cr}.
Overall, the worst-case computational complexity can be approximated as $\mathcal{O} \left (  \log{ \left ( \zeta^{-1} \right ) } \left [ KN^2  \right ]^{3.5} \right )$.

\section{NUMERICAL RESULTS}
\begin{figure}[ht]
      \centering
	\includegraphics[width=8.11cm]{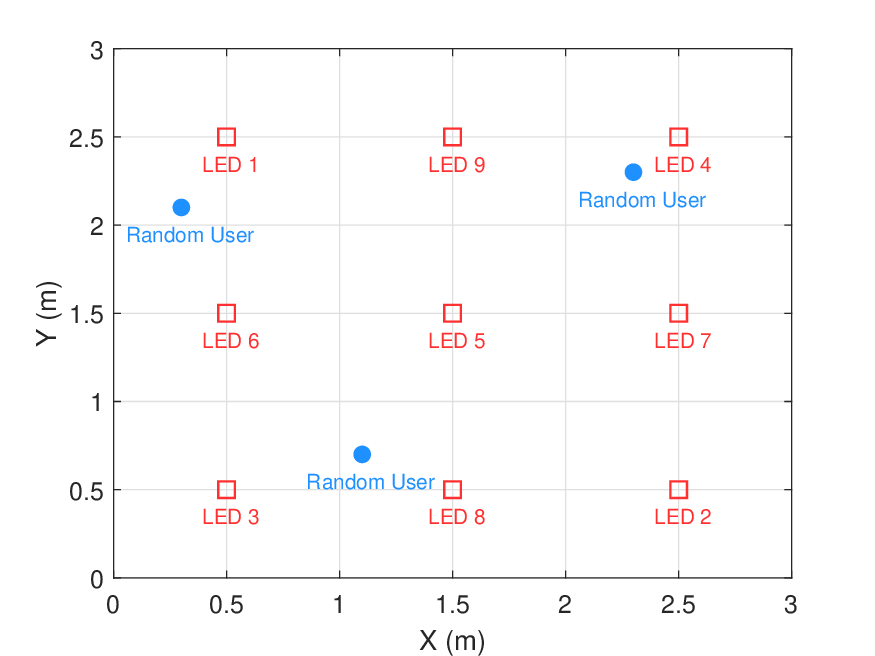}
\caption{Room configuration with LEDs’ locations.}
  \label{room}
\end{figure}

The final section here aims at evaluating the proposed robust beamforming design of the RSMA-aided VLC network with imperfect CSIT. 
We consider a downlink multi-user VLC network deployed in a certain room with a fixed size of $3$ m $\times$ $3$ m $\times$ $5$ m. 
A three-dimensional coordinate $(X, Y, Z)$ is applied to model the locations of both LEDs and users in fixed room.
The origin $(0, 0, 0)$ is a corner of this room.
There exists in total nine LEDs $(N=9)$ in the ceiling serving multiple users uniformly distributed in the receiving plane. 
The horizontal distance between the LEDs and users is fixed to $1.7$ m, i.e., the user location for user-$k$ is defined as ($x_k$, $y_k$, $1.7$).
Without loss of generality, the peak amplitude of signal, the variance of input signal, the average electrical noise power, and the radius of uncertainty region are equal for all users, i.e., $A = A_0 = A_1 = \cdots  = A_K,\ \varepsilon = \varepsilon_0   = \varepsilon_1 = \cdots  = \varepsilon_K$, $\sigma^2 = \sigma^2_1 = \cdots  = \sigma^2_K$, and $r = r_1 = \cdots  = r_K$.
Room configuration details are displayed in Fig. \ref{room}.
The locations of both LEDs and users are specified in Table \ref{table_1}.
In Table \ref{table_2}, we further illustrate detailed parameters of the considered VLC network.

\begin{table}[ht]
\centering
\caption{\normalsize LOCATIONS OF LEDS.}
\begin{tabular}{|c|c|c|c|}
\hline
       & Coordinate &        & Coordinate  \\

\hline
$U_k$  &($x_k$,$y_k$,1.7)  &${\mathrm{LED }}\;1$ & (0.5,2.5,4.5)\\
  \hline
${\mathrm{LED }}\;2$ & (2.5,0.5,4.5)& ${\mathrm{LED }}\;3$ &(0.5,0.5,4.5)\\
  \hline
${\mathrm{LED }}\;4$ &(2.5,2.5,4.5) & ${\mathrm{LED }}\;5$ &  (1.5,1.5,4.5)\\
  \hline
${\mathrm{LED }}\;6$ &(0.5,1.5,4.5)&${\mathrm{LED }}\;7$ &(2.5,1.5,4.5)\\
  \hline
 ${\mathrm{LED }}\;8$  &  (1.5,0.5,4.5) & ${\mathrm{LED }}\;9$ &(1.5,2.5,4.5)  \\
  \hline
\end{tabular}\label{table_1}
\end{table}

\begin{table}[htbp]
	\caption{SUMMARY OF LED PARAMETERS AND THEIR VALUES. }
	\label{table_2}
	\centering
	\begin{tabular}{|m{3.5cm}|m{1cm}<{\centering}|m{2.2cm}<{\centering}|}
        \hline
		Peak amplitude of signals  & $A$ & $2\;{\mathrm{V}}$\\
		\hline
		Variance of signals & $\varepsilon$  & 1 \\
        \hline
		Receiver FOV & ${\varphi _{{\mathrm{FOV}}}}$ & ${60^ \circ }$ \\
		\hline
		LED emission semi-angle & ${\phi _{1/2}}$ & ${60^ \circ }$ \\
       \hline
		PD geometrical area & ${A_{{\mathrm{PD}}}}$ & $1\;{\mathrm{c}}{{\mathrm{m}}^2}$ \\
		\hline
		Refractive index & $n_r$ & $1.5$ \\
		\hline
		PD responsivity  & ${\theta _l}$ & ${\mathrm{0}}{\mathrm{.54}}\left( {{\mathrm{A/W}}} \right)$\\
		\hline
		Average electrical noise power  & ${\sigma}^2$ & $-98.82\;\mathrm{dBm}$\\
		\hline
		DC bias  & $b$ & $\sqrt 6 \;  $\\
		\hline
	\end{tabular}
\end{table}
To evaluate the performance of the proposed robust beamforming design algorithm for RSMA-aided VLC networks, we consider NOMA and SDMA-assisted VLC networks as two benchmark schemes. 
Readers are referred to \cite{mashuai_mobile} and \cite{S_Ma_2017} for more details of the baselines. 
It is worth noting that the proposed algorithm can be directly applied to address the MMF problem for both SDMA and NOMA-aided VLC networks since SDMA and NOMA are two special instances of RSMA.
In order to make a comprehensive comparison among RSMA, NOMA and SDMA in VLC networks, both network loads, specifically underloaded and overloaded regimes are taken into account. \par
\begin{figure}[ht]
      \centering
	\includegraphics[width=8.11cm]{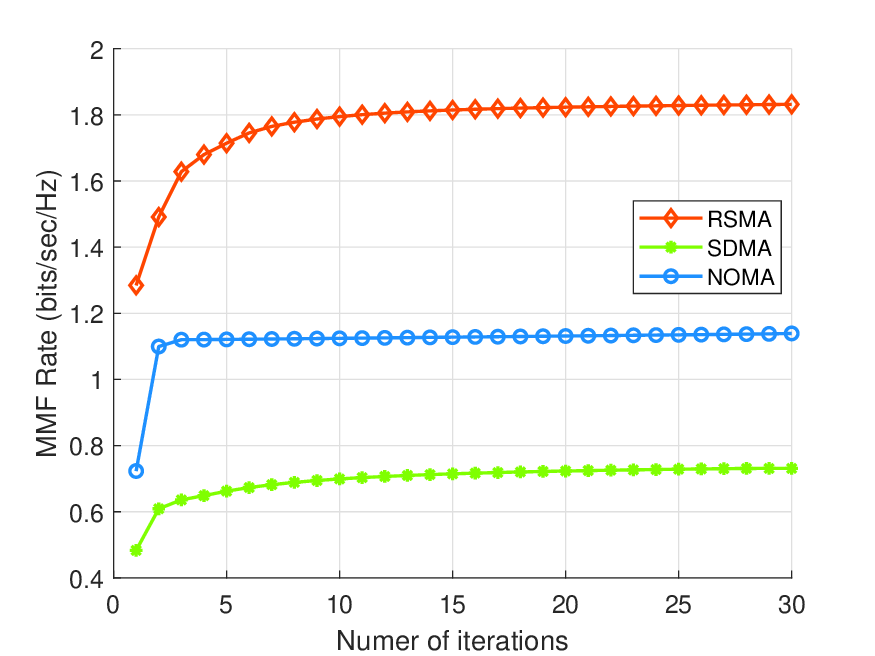}
\caption{Convergence analysis of the proposed algorithm.}
  \label{convergence}
\end{figure}
We first assess the convergence performance of the proposed algorithm. 
Specifically, we consider a VLC BS with 4 LEDs, i.e., LED 1,2,3 and 4 as listed in Table \ref{table_1}.
They serve four users with SNR = 15 dB, $\left [ I_L,I_H \right ]  = \left [ 15,20 \right ]$ mA, and a uncertainty radius of $r=0.05$ m. 
All results in the following consider the same uncertainty radius, unless otherwise noted.
Fig. \ref{convergence} illustrate the convergence behaviour of the proposed algorithm for one randomly and uniformly generated user location.
It is obvious that the proposed algorithm can converge within few iterations.
\begin{figure}[ht]
      \centering
	\includegraphics[width=8.11cm]{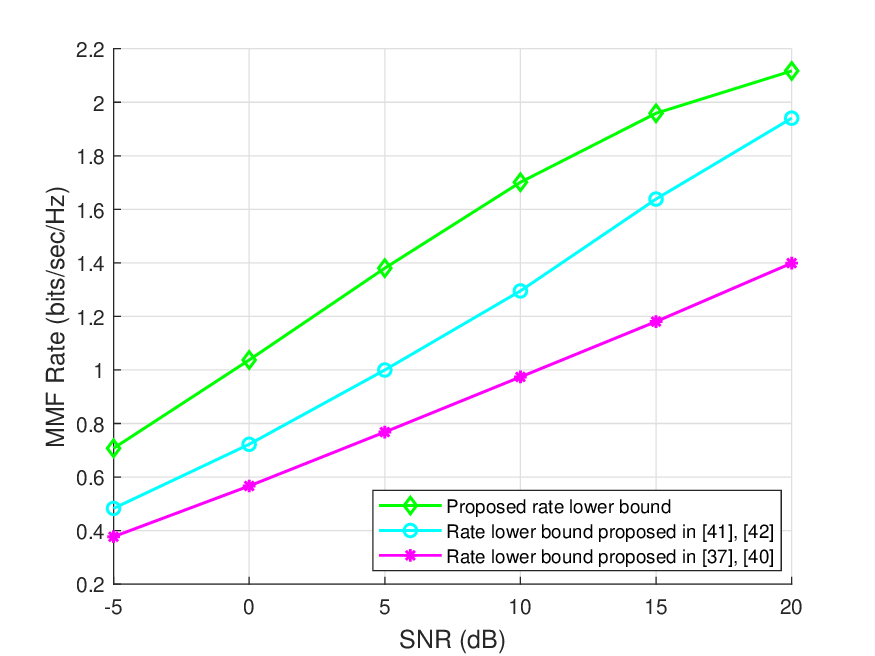}
\caption{The MMF rate (bits/sec/Hz) versus SNR (dB) under different lower bounds of rate expressions where 4 LEDs serve 4 users.}
  \label{dif_rate}
\end{figure}
\par
We first compare our derived rate lower bound with conventional rate lower bounds used in some existing works \cite{Xing_2022,2015_pham,2024_guo,2013_wang,2009_lapidoth}. 
Commonly used rate lower bounds typically take the form $\frac{1}{2} \log_2(1+\varpi\mathrm{SINR})$, where the coefficient $\varpi$ is predominantly determined by the input distributions.
The most common values for $\varpi$ is $\frac{2}{\pi \mathrm{e}}$ or $\frac{\mathrm{e}}{2 \pi }$, depending on whether the input distribution follows a uniformly spaced discrete distribution, as in \cite{Xing_2022,2015_pham}, or a particular closed-form distribution, as in \cite{2024_guo,2013_wang,2009_lapidoth}.
As illustrated in Fig. \ref{dif_rate},  
the MMF rate for our derived distribution in \eqref{abg_distribution} is  higher than the conventional rate lower bounds. Notably, our derived rate lower bound accounts for residual interference resulting from imperfect SIC, which constrains its performance in high-SNR regimes. Nevertheless, the proposed rate lower bound still achieves a non-negligible rate gain over the two baselines in this regime.

\begin{figure}[ht]
      \centering
	\includegraphics[width=8.11cm]{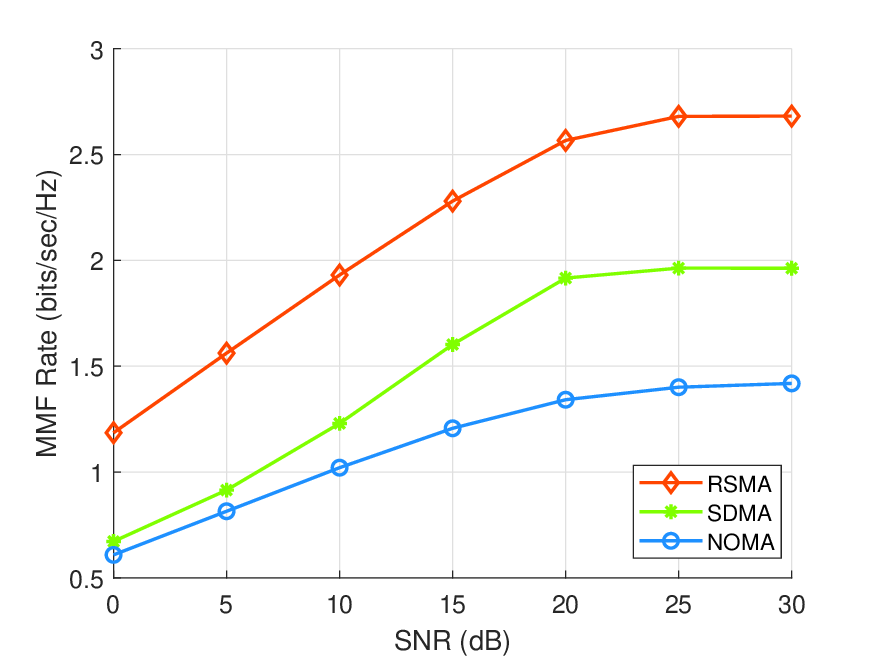}
\caption{The MMF rate (bits/sec/Hz) versus SNR (dB) for an underloaded regime where 6 LEDs serve 4 users.}
  \label{6s4}
\end{figure}

\begin{figure}[ht]
      \centering
	\includegraphics[width=8.11cm]{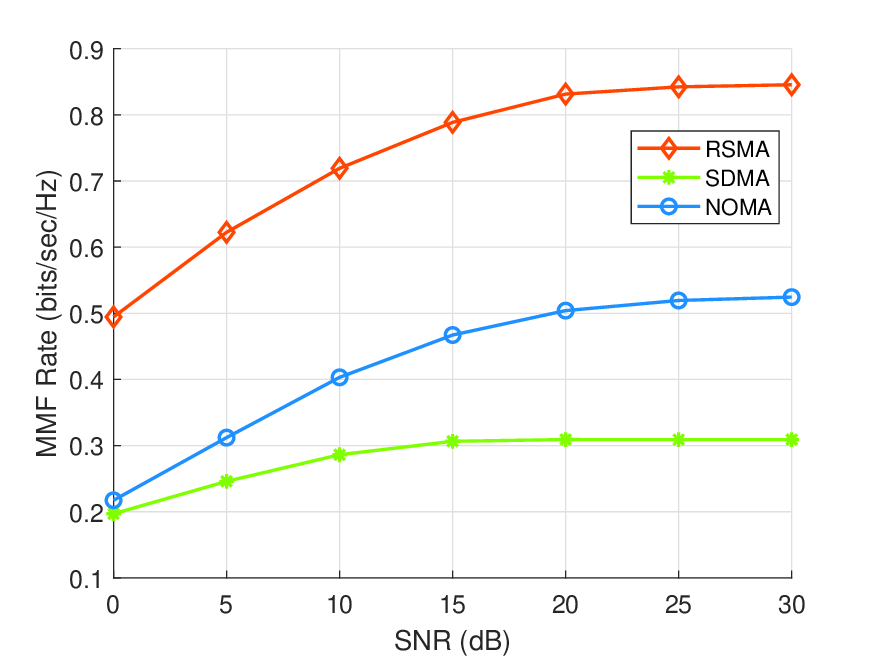}
\caption{The MMF rate (bits/sec/Hz) versus SNR (dB) for an overloaded regime where 6 LEDs serve 8 users.}
  \label{6s8}
\end{figure}

Fig. \ref{6s4} and Fig. \ref{6s8} illustrates the MMF rate (bits/sec/Hz) versus SNR (dB) for different schemes in both underloaded and overloaded regimes when $N=6$ LEDs serve $K=4$ users or $K=8$ users.
Here the location of the 6 LEDs corresponds to LED 1,2,3,4,6, and 7 as listed in Table \ref{table_1}.
It can be observed from both figures that the MMF rates of all three schemes increase with SNR, and RSMA always surpasses NOMA and SDMA in both underloaded and overloaded regimes. 
Specifically, RSMA attains nearly $150 \%$ MMF rate gain over SDMA and $185 \%$ MMF rate gain over NOMA in the underaloded regime. 
As for the overloaded regime, RSMA achieves approximately $274 \% $ MMF rate gain compared to SDMA and $180 \%$ MMF rate gain compared to NOMA.
This is because RSMA has the ability of partially decoding the interference and partially treating the interference as noise through message splitting at the transmitter and message combining at the receivers. 
Besides, we observe that the MMF rate saturates in the high SNR regime. 
This is due to the additional optical power limitation of each LED imposed by constraint \eqref{constraint_ini_3}. 
As both the optical power limitation and the electric power constraint are considered, the overall MMF rate performance is limited by the optimal power constraint when the electric power constraint is large in the high SNR regime. \par

\begin{figure}[ht]
      \centering
	\includegraphics[width=8.11cm]{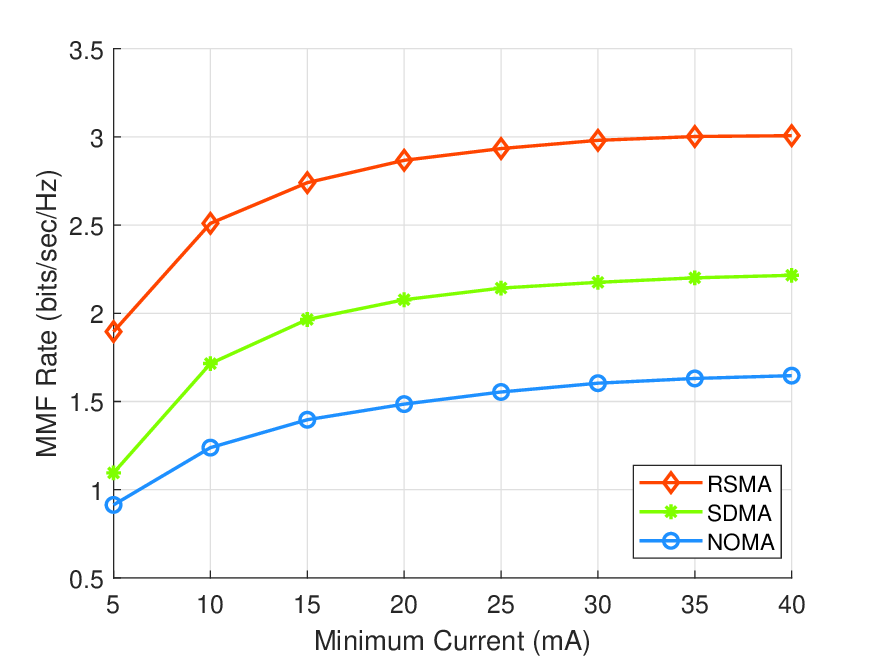}
\caption{The MMF rate (bits/sec/Hz) versus the Minimum current $I_L$ (mA) for an underloaded regime where 6 LEDs serve 4 users.}
  \label{6s4_mmf_vs_dc}
\end{figure}

To further explore the influence of the optical power constraint, we investigate the MMF rate (bits/sec/Hz) versus the minimum current $I_L$ (mA) of different schemes in Fig. \ref{6s4_mmf_vs_dc} when SNR = 30 dB and $I_H = I_L + 5$ (mA).
We observe that the MMF rate of three strategies increases with the minimum current until constraint \eqref{constraint_ini_4} dominates again in the two power constraints. 
For all different power constraints, RSMA still outperforms other two strategies.

\begin{figure}[ht]
      \centering
	\includegraphics[width=8.11cm]{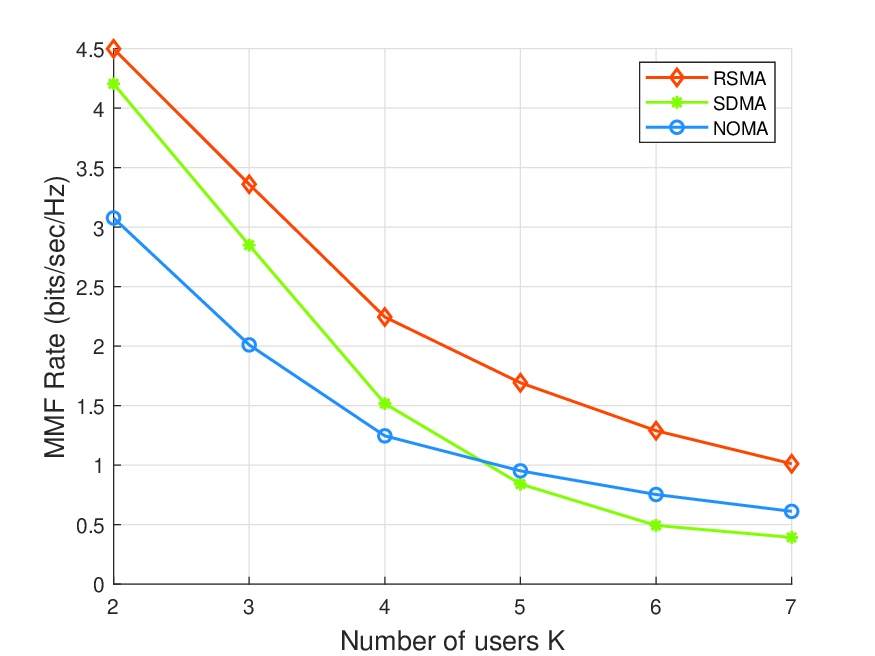}
\caption{The MMF rate (bits/sec/Hz) versus the number of users $K$ when $N=6$.}
  \label{mmf_vs_user_num}
\end{figure}

\begin{figure}[ht]
      \centering
	\includegraphics[width=8.11cm]{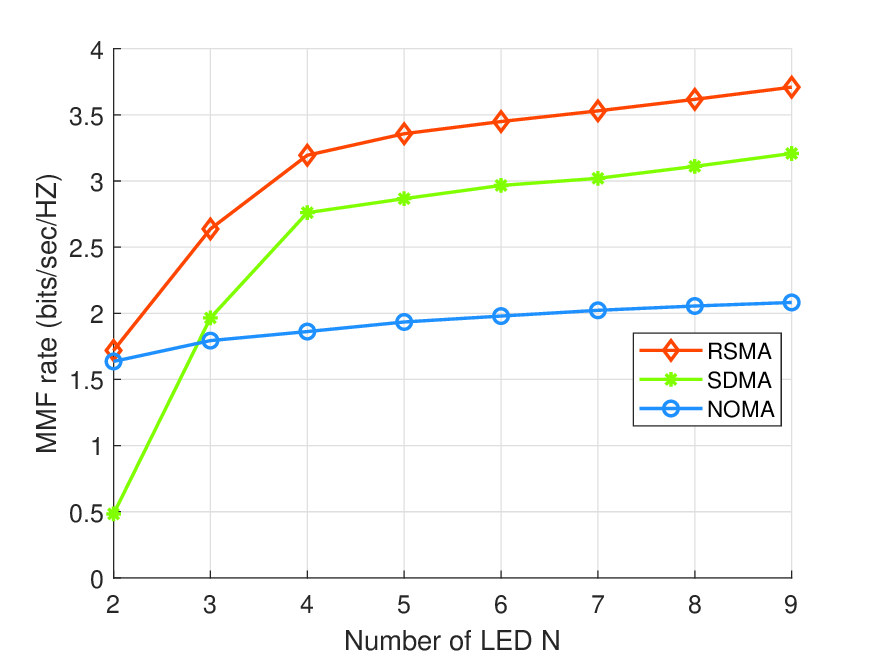}
\caption{The MMF rate (bits/sec/Hz) versus the number of LEDs $N$ when SNR = 15 dB and $K = 3$.}
  \label{2-9s3}
\end{figure}
Moreover, we observe that the SDMA scheme performs better than the NOMA scheme in the underloaded regime where $N=6$  LEDs serve $K=4$ users.
However, as specified in Fig. \ref{6s8}, SDMA performs worse than NOMA in the overloaded regime when the number of users exceeds the number of LEDs. 
This is due to the severe multi-user interfernece, which degrades the performance of SDMA.

Furthermore, we investigate the MMF rate (bits/sec/Hz) versus the number of users $K$ in Fig. \ref{mmf_vs_user_num} when $N=6$ and SNR = 15 dB.
We observe that the MMF rate of all three strategies decreases with more users engaged, while the performance of SDMA drops significantly as the number of users increases.
This is because SDMA is highly sensitive to the network load and user deployment and can not manage effectively stronger interference as the number of users increases.
Specifically, RSMA attains nearly $107 \%$ MMF rate gain over SDMA and $146 \%$ MMF rate gain over NOMA in a highly underloaded network where $6$ LEDs only serve $2$ users. 
In an overloaded network load setting, where $6$ LEDs serve $7$ users, the advantages of RSMA become more pronounced. RSMA achieves an MMF rate gain of approximately $257 \%$ over SDMA and $165 \%$ over NOMA.

We further illustrate the MMF rate versus the number of LEDs with SNR = 15 dB and fixed number of users $(K = 3)$ in Fig. \ref{2-9s3}. 
We observe that the MMF rate of all three schemes increases with the number of LEDs. 
This is because multiple LEDs at the VLC transmitter provide a higher flexibility in beamforming design to manage interference. 
SDMA and NOMA outperforms each other in the underloaded and overloaded regimes, while RSMA always outperforms these two baseline schemes in all regimes.

\begin{figure}[ht]
      \centering
	\includegraphics[width=8.11cm]{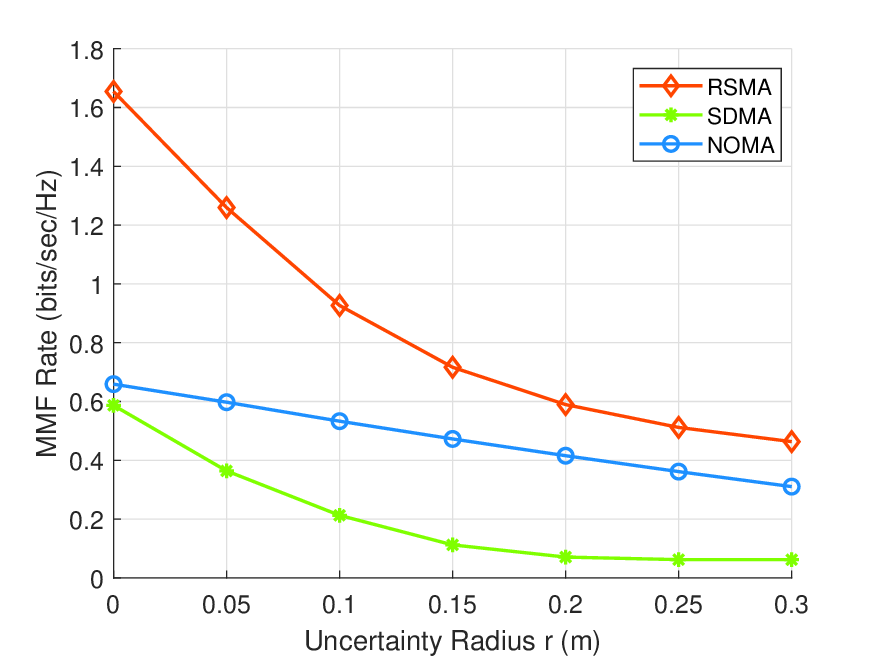}
\caption{The MMF rate (bits/sec/Hz) versus the CSI uncertainty radius $r$ for RSMA, SDMA and NOMA schemes when SNR = 10 dB, $N=6$ and $K = 6$.}
  \label{6s6_uncertainty_change}
\end{figure}
In order to explore the influence of user mobility radius, we plot the MMF rate versus the uncertainty region radius $r$ of mobile users in Fig. \ref{6s6_uncertainty_change}, when $N=6$, $K=6$, and SNR = 10 dB.
It can be observed that as the radius increases, the performance of all three strategies decreases. 
RSMA scheme always surpasses both SDMA and NOMA scheme in all practical uncertainty radius.
Specifically, under a small uncertainty radius, i.e., $r=0.05$ m, the MMF rate of RSMA surpasses NOMA scheme by nearly $210 \%$.
When it comes to a larger uncertainty radius, i.e., $r=0.3$ m, RSMA surpasses NOMA scheme by nearly $149 \%$.
It is obvious that SDMA achieves marginal MMF rate when the uncertainty region radius is large.
This is due to the fact that, when the uncertainty region is large, CSIT becomes inaccurate.
SDMA cannot effectively manage multi-user interference with inaccurate beamforming.

\section{CONCLUSION}
In this paper, we consider a RSMA-aided VLC network with imperfect CSI and propose a robust beamforming design to address the non-convex MMF problem. 
Specifically, we first derive the theoretical lower bound of the VLC channel capacity. 
Based on the close-form expressions, we propose a robust beamforming design method to maximize the worst-case rate among users with practical imperfect CSI. 
Specifically, we apply SDR, $\mathcal{S}$-lemma, CCCP, and propose a penalty-based method to acquire high-quality and sub-optimal beamformers. 
Numerical results show that the proposed RSMA-aid VLC beamforming design scheme surpasses the conventional SDMA and NOMA schemes.
Our future research directions include exploring practical VLC scenarios, such as incorporating non-line-of-sight (NLOS) reflection and diffuse paths. Additionally, we aim to extend RSMA-aided VLC to support simultaneous lightwave information and power transfer (SLIPT) as well as integrated sensing, lighting, and communications.

\bibliographystyle{ieeetr}
\bibliography{VLC_RSMA_manuscript}

\end{document}